\documentclass[11pt]{article}

\usepackage[final]{acl}

\usepackage{times}
\usepackage{latexsym}

\usepackage[T1]{fontenc}

\usepackage{hyperref}
\usepackage{url}
\usepackage{colortbl}
\usepackage{nicefrac}       %
\usepackage{microtype}      %
\usepackage[dvipsnames]{xcolor}         %
\usepackage{multirow}
\usepackage{multicol}
\usepackage{graphicx}
\usepackage{xspace}
\usepackage{amsmath}
\usepackage{adjustbox}
\usepackage{tcolorbox}
\usepackage{amssymb}
\usepackage{enumitem}
\usepackage{wrapfig}
\usepackage{xcolor}
\usepackage{subcaption}    %
\usepackage{float}
\usepackage{stfloats}
\usepackage{amsmath}
\usepackage{listings}
\usepackage{mdframed}
\usepackage{pifont}
\usepackage{tcolorbox}
\usepackage{arydshln}
\usepackage{booktabs} 
\usepackage{color,soul}
\usepackage{makecell}
\usepackage{fontawesome5}

\usepackage{cleveref}
\usepackage{enumitem}
\usepackage{tabularx}
\usepackage[table]{xcolor}

\tcbuselibrary{listings,breakable}

\definecolor{pastelblue}{RGB}{173,216,230}
\definecolor{pastelyellow}{RGB}{255,253,208}
\definecolor{pastelpink}{RGB}{255,209,220}
\definecolor{pastelgreen}{RGB}{176,226,172}
\definecolor{pastellavender}{RGB}{230,230,250}

\definecolor{NvidiaGreen}{RGB}{118, 185, 0}
\sethlcolor{red!15}

\newcommand{\thinking}{{ \includegraphics[height=1.1em]{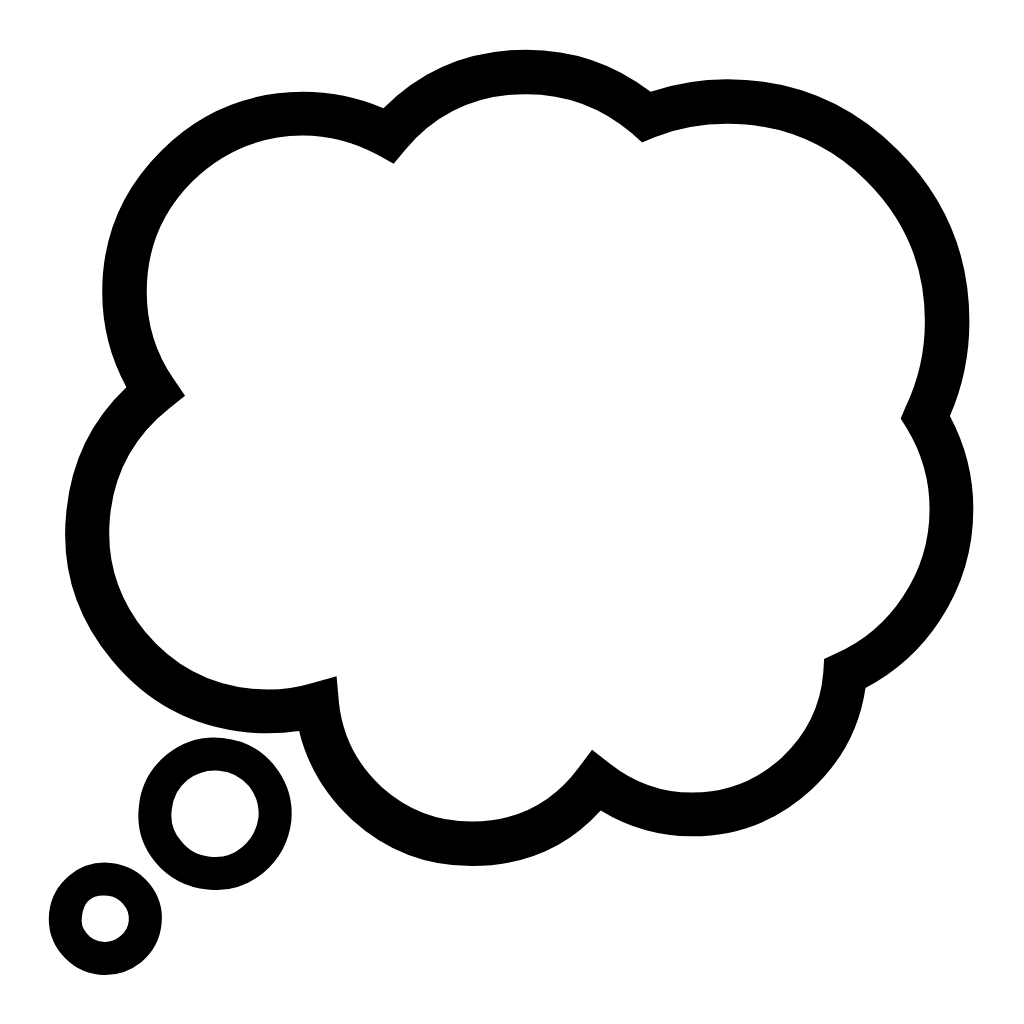}}}
\newcommand{\moe}{{ \includegraphics[height=1.1em]{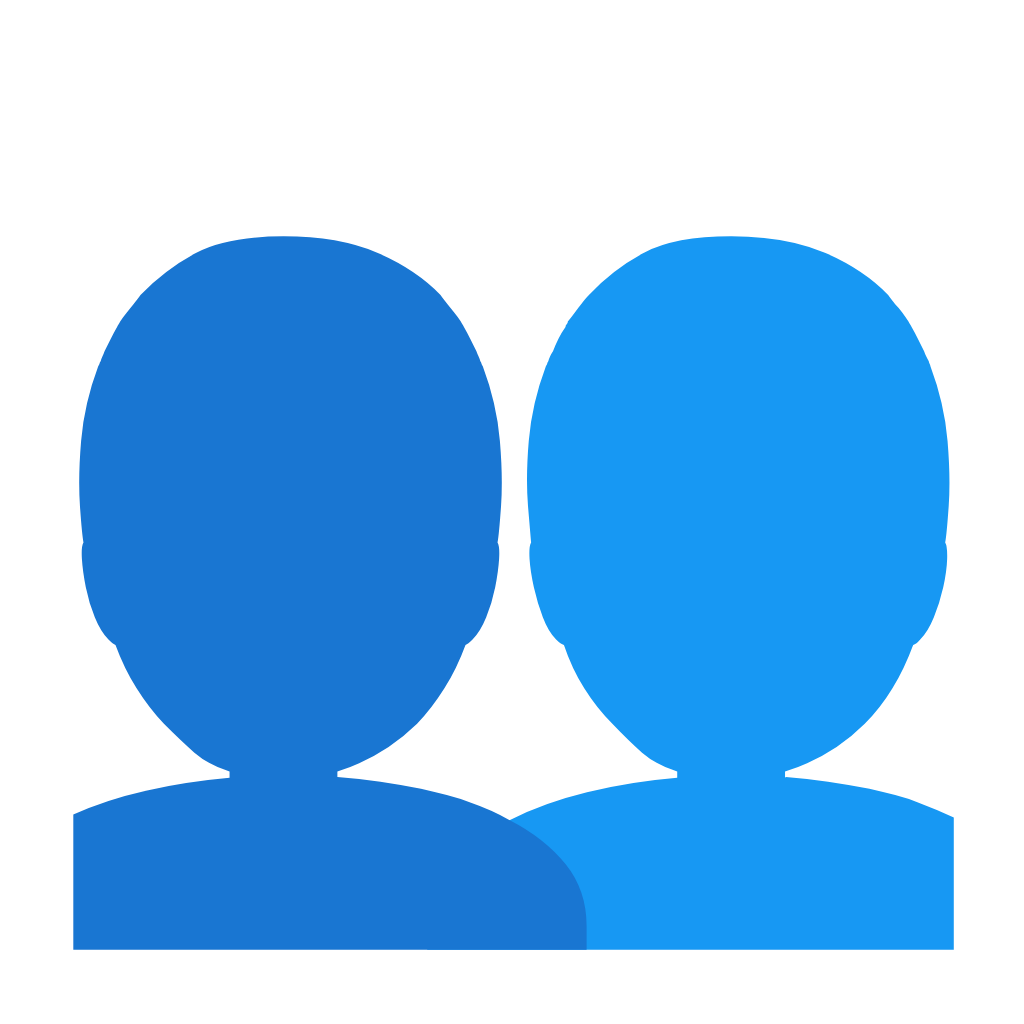}}}
\newcommand{\science}{{\includegraphics[height=1.1em]{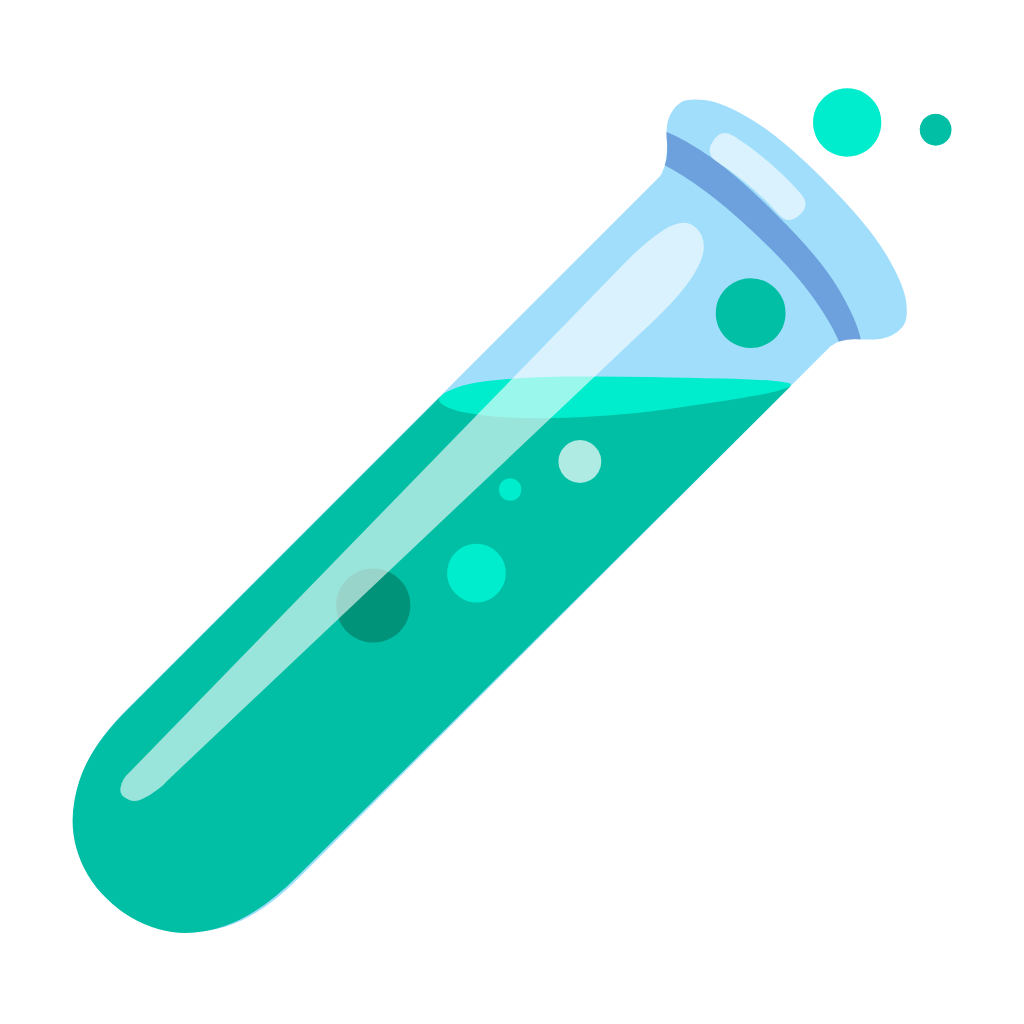}}}
\newcommand{\llama}{{\includegraphics[height=1.1em]{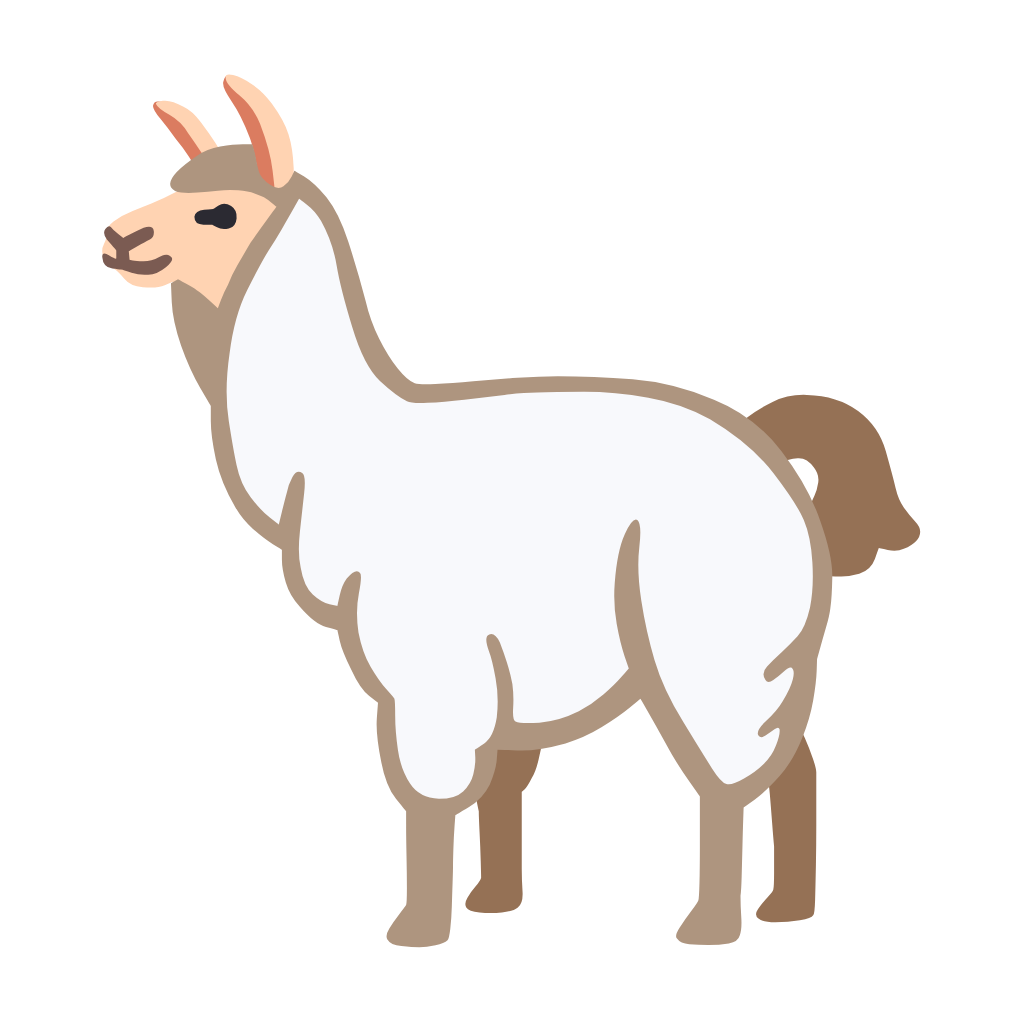}}}
\newcommand{\qwen}{{\includegraphics[height=1.1em]{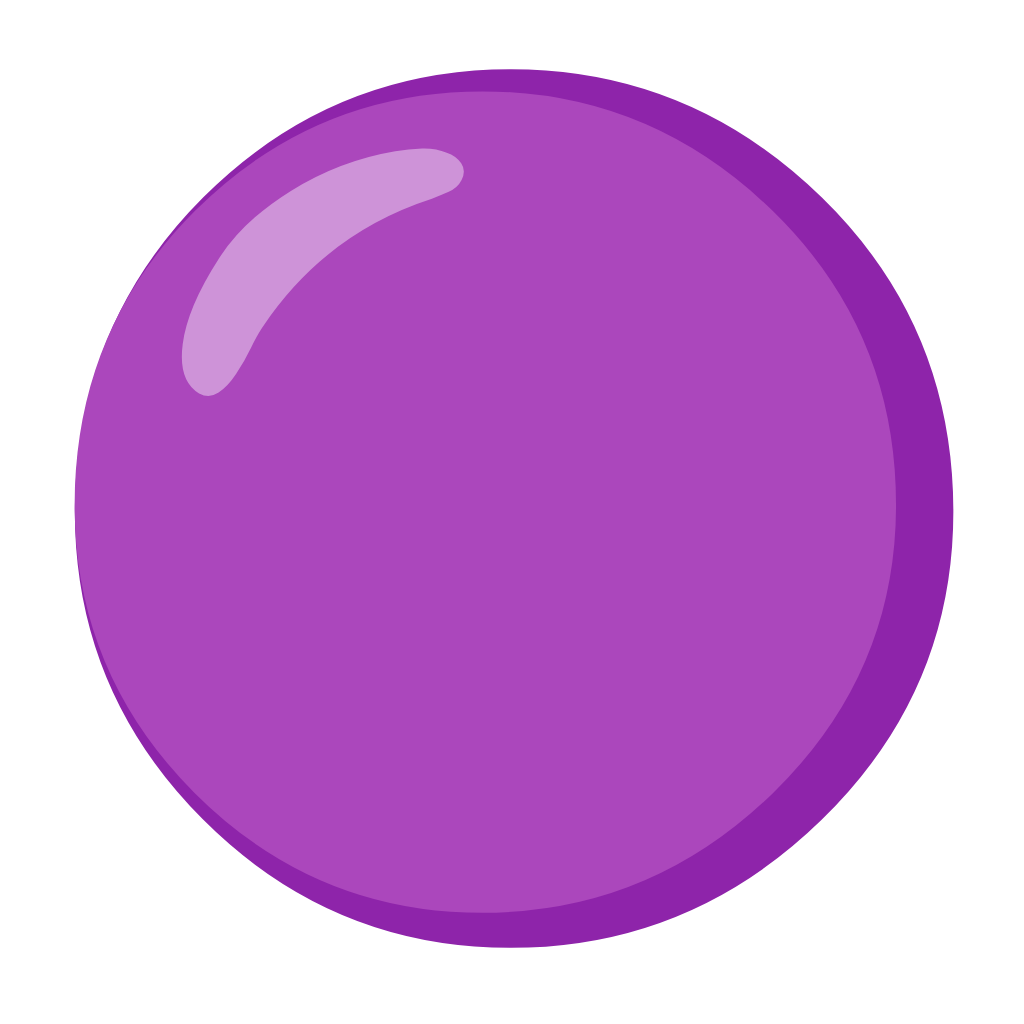}}}
\newcommand{\AxE}{\textcolor{magenta}{$\Large\mathord{\text{\ding{58}}}$}}
\newcommand{\AxI}{\textcolor{Aquamarine}{$\Large\mathord{\text{\ding{54}}}$}}
\newcommand{\Err}{\includegraphics[height=1.1em]{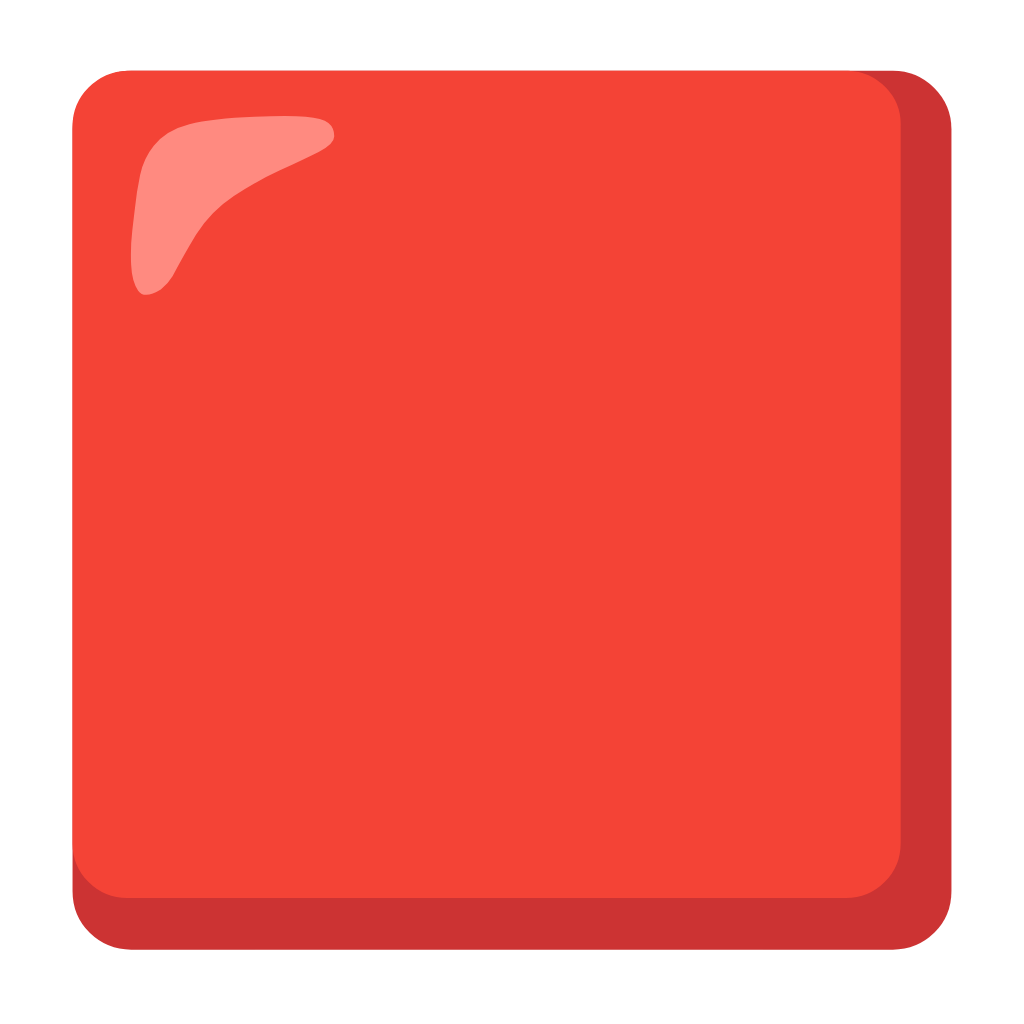}}
\newcommand{\IoU}{\textcolor{Green}{\LARGE$\blacktriangle$}}
\newcommand{\Acc}{\includegraphics[height=1.1em]{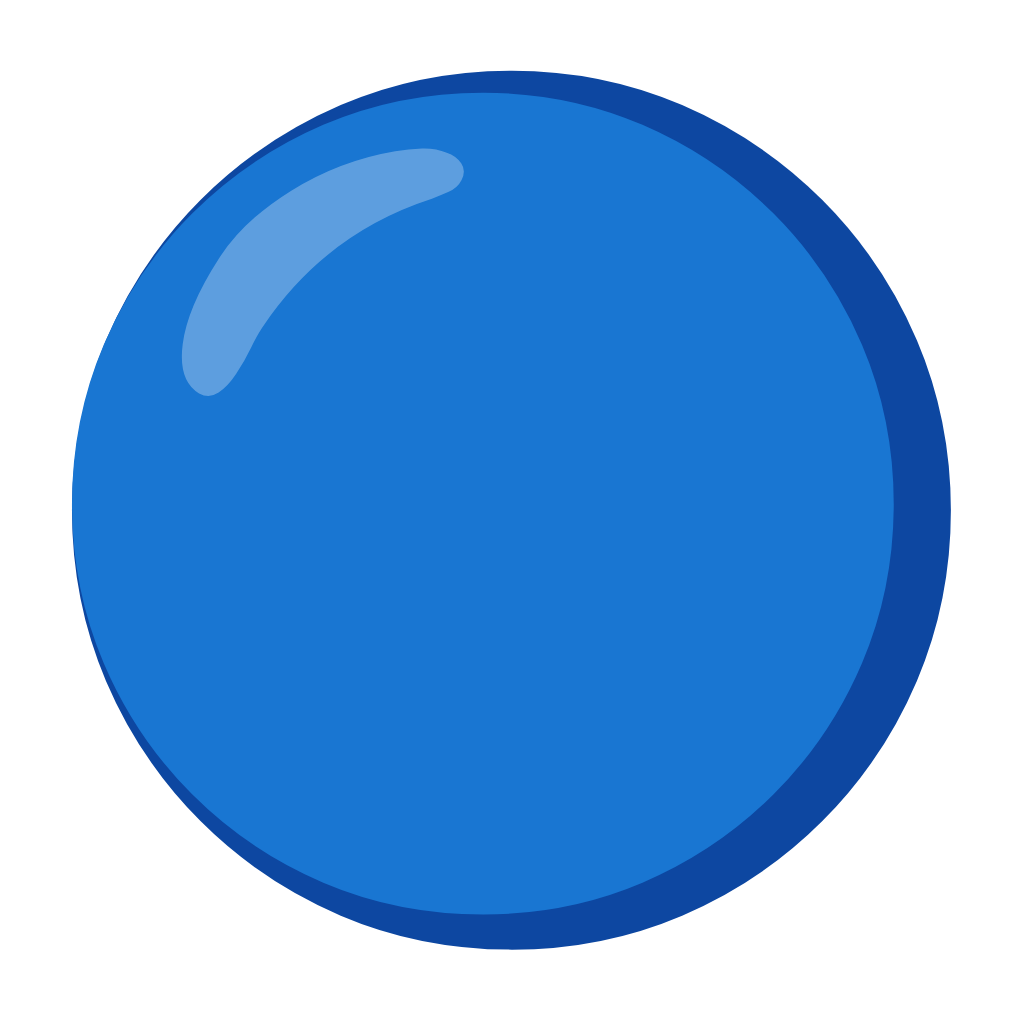}}
\newcommand{\Siz}{\includegraphics[height=1.1em]{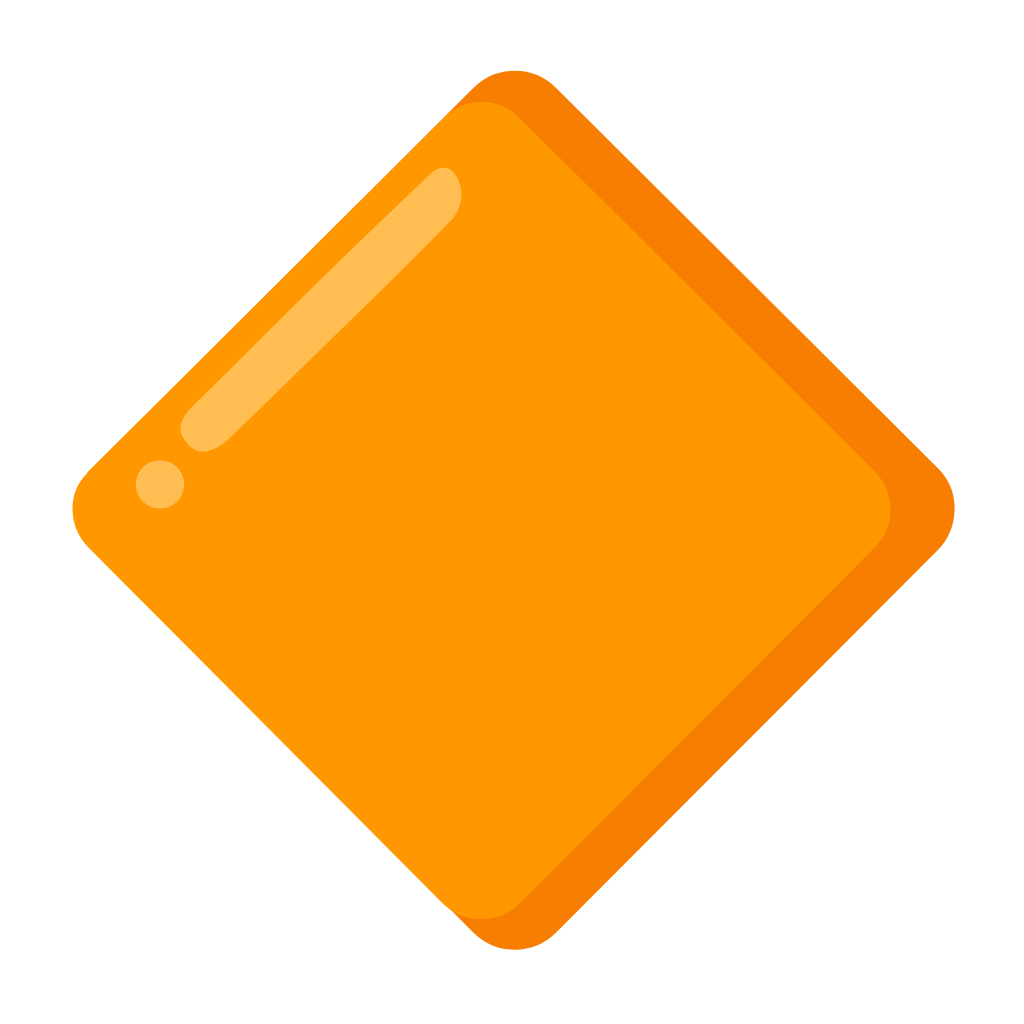}}
\newcommand{\Llm}{\textcolor{Brown}{\LARGE$\blacktriangledown$}}
\newcommand{\Fam}{\includegraphics[height=1.1em]{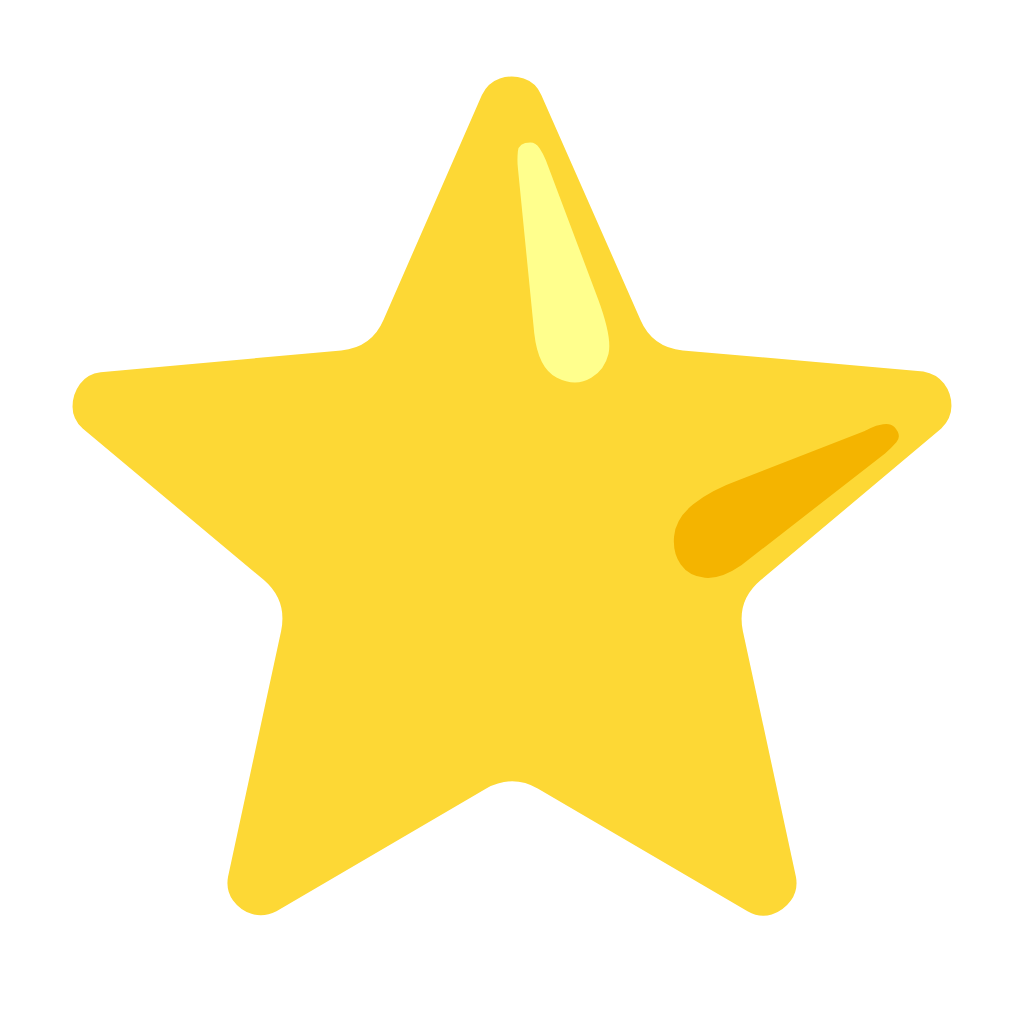}}

\title{\centering{Mo' Models, Mo' Problems: How to best select model pools when designing Multi-Agent Systems}}

\author{
 \textbf{Sara Vera Marjanović\textsuperscript{1*}},
 \textbf{Jiacheng Xu\textsuperscript{2}},
 \textbf{Aleksandr Laptev\textsuperscript{2}},\\
 \textbf{Grigor Nalbandyan\textsuperscript{2}},
 \textbf{Erik Arakelyan\textsuperscript{2}},
 \textbf{Evelina Bakhaturina\textsuperscript{2}},
\\
 \textsuperscript{1}University of Copenhagen
 \textsuperscript{2}NVIDIA
\\
 \textsuperscript{*}Work done during internship at NVIDIA
\\
 \small{
   \textbf{Correspondence:} \href{mailto:jiachengx@nvidia.com}{jiachengx@nvidia.com}
 }
}

\begin{document}
\maketitle
\begin{abstract}
Multi-agent Systems (MAS) combine multiple model outputs to solve complex reasoning tasks. However, despite rapid growth of available open-source models, there is limited research on how to select optimal model candidates out of this massive pool. We systematically evaluate 8 model selection strategies (including model size, accuracy and answer diversity) across before-generation (routing) and after-generation (majority-voting, LLM-as-a-judge) MAS architectures on challenging scientific benchmarks. Our findings show a significant gap between theoretical oracle potential and actual performance: Expanding candidate pool sizes often degrades performance below that of the top performing base-model. We find that candidate selection within a single model family is the strategy that yields the best relative performance over a standalone model. These results demonstrate that adding arbitrary models to a heterogeneous MAS can introduce system instability, highlighting model selection as a critical design choice for multi-agent systems.
\end{abstract}

\maketitle

\hspace{3em}
\faGithub 
\hspace{2mm}\href{https://github.com/spaidataiga/mo-models}{spaidataiga/mo-models} %
\vspace{0.5em}

\section{Introduction}

\begin{figure*}
    \centering
    \includegraphics[width=\linewidth]{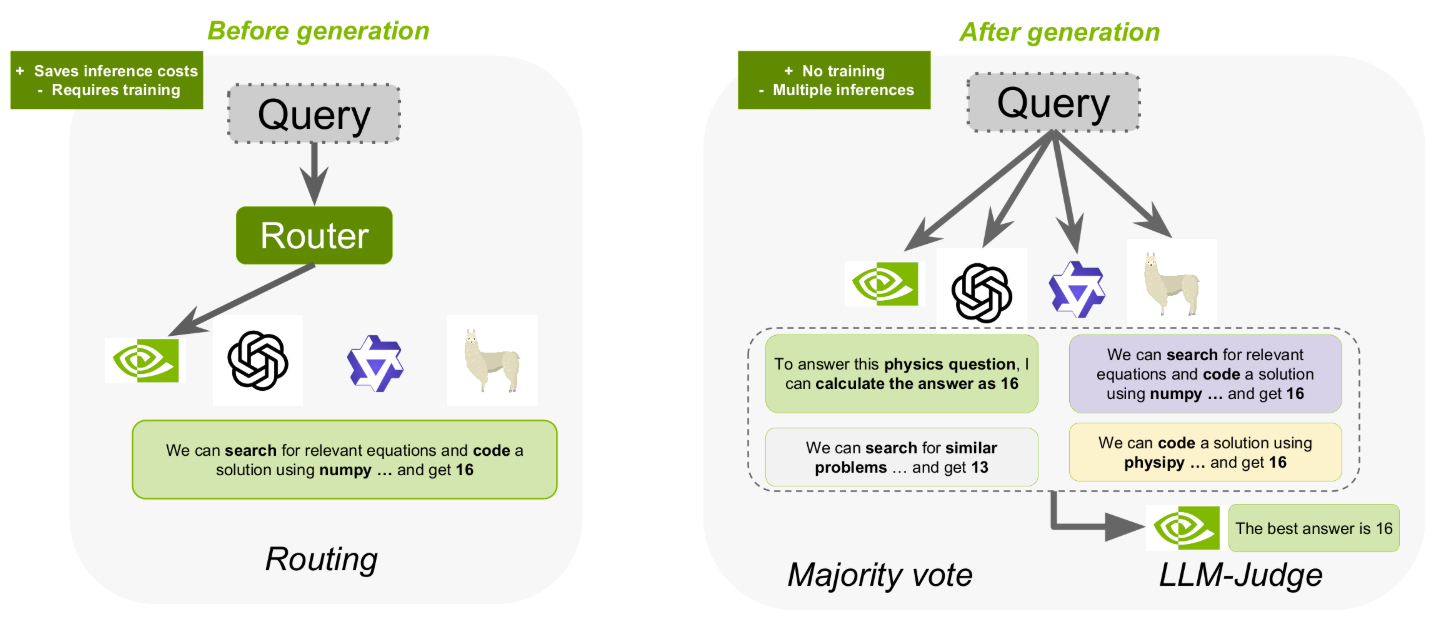}
    \caption{Before and After-generation Multi-Agent Systems}
    \label{fig:mas_types}
\end{figure*}

As modern large language models (LLMs) advance, progress is measured on challenging benchmarks designed outside the limits of current state-of-the-art models (HLE \cite{hle}, BIG-Bench Extra Hard \cite{bigbenchxh}), motivating new system-level approaches. Recently multi-agent frameworks like  Kimi-K2.5 Agent Swarm \cite{agentswarm}, and Virtual AI Lab \cite{virtuallab} have shown success in pushing model capabilities. By using multiple instances of the same base model to collaborate on a query, these ensembles leverage multiple reasoning strategies and perspectives, enabling stronger performance on increasingly challenging reasoning problems \cite{pmlr-v235-du24e,kim2026reasoning}.  %

Carefully combining different models could yield further gains than systems built from one base model \cite{ICLR2025_mixtureofagents}. Generalist models often have their own emergent specializations \cite{pmlr-v235-chiang24-chatbotarena, liu-etal-2024-mathbench,srivastava2023beyond}, and even small specialist models can outperform large generalists on targeted tasks \cite{chemdfm,chemr,cosmosage}, also reducing costs. Satisfactorily combining heterogeneous models into a successful multi-agent system (MAS) must somehow collapse multiple diverse generations into one final answer. While there are many possible MAS structures \cite{chen2026llmensemblesurvey, masgpt_pmlr2025}, there are two basic forms: \textit{before-generation MAS}, where a router determines the ideal model(s) to answer a query, and \textit{after-generation MAS}, where multiple answers are generated and collapsed into one final response (See \Cref{fig:mas_types,sec:background}). More complex MAS combine these systems \cite{su2026toolorchestra, ICLR2025_mixtureofagents}. 

A seemingly overlooked MAS design choice is the \textit{selection of agents}. Given a space of nearing infinite models available on huggingface\footnote{As of July 2026, there were 2,914,945 models available on huggingface. In the two months taken to review this paper, over 150k more models were uploaded.} $\mathcal{M}=\{m_1,m_2,...,m_\infty\}$, is it always better to select as many as computationally possible? Given information on a model card, or additional model evaluation, how can we best select our subset of candidate models $C\subset \mathcal{M}$ to optimise both performance and costs? 

In this work, we explicitly evaluate the effect of model candidate selection for MAS, spanning both before-generation and after-generation paradigms.  We compare selections of $C$ as determined by performance, answer diversity, or architecture, and evaluate these strategies in terms of their potential and actual performance on challenging science reasoning tasks.

We find that MAS model performance cannot be predicted without first evaluating candidate models: domain specialisations are not always realised, and different architectural families show different levels of similarity. High oracle performance does not necessarily provide actual improvements in MAS performance; oftentimes, systems with high prophesised improvement do not perform better than the best base model. While homogeneous MAS improve with increased architectural complexity, model instances and diversity, we do not observe this with our heterogeneous set-up: the few times a MAS improves over a single  base model is when the elements of our MAS come from one singular family of models. Therefore, we advise creators of MAS architectures to evaluate and report \textit{which} model behaviour patterns are required for success of their systems.

\section{Related Work}\label{sec:background}

\paragraph{Non-LLM Ensembles} Ensembling methods have long been a cornerstone of machine learning to increase accuracy and robustness \cite{ensembleml}. Typically, individual models within the ensemble are intentionally trained to be diverse, though the representation of diversity differs: for example, one model architecture can be trained multiple times on \textit{different subsets of data} to increase robustness \cite{bagging_ensemble,randomforests}. In model stacking, models of \textit{different architectures} trained on a shared training dataset are used to create a final model \cite{modelstacking}. Model stacking requires that the original trained model have high \textit{error diversity}, motivating the use of varying architectures. However, error diversity can also be promoted in an ensemble via negative correlation learning \cite{ncl1999_ensemble} and PAC-Bayes C-bound \cite{NIPS2006_pacbayes}. Conversely, \textit{correct-answer diversity} in an ensemble system can be encouraged via gating networks, allowing the creation of complementary experts for routing in Mixture-of-Experts systems \cite{moe_ensemble}.

\paragraph{Before-generation MAS} Routing MAS learn to map queries to an optimal, \emph{frozen}, model $m$ in $\mathcal{M}$. This may be done to balance complementary skills \cite{wang2024benchcoe,ICLR2024_frugalfoe}, reduce inference costs \cite{Varangot-routing_survey, ong2025routellm}, and/or provide parallelization in complex MAS \cite{Varangot-routing_survey, wang2024benchcoe, shao2025routeandreasonscalinglargelanguage, ICLR2025_mixtureofagents}. The router backbone can range in complexity from simple functions \cite{Zhang_2025avengers, hari2023tryage} to finetuned language models \cite{su2026toolorchestra, ong2025routellm, mohammadshahi2024routoo}; it can route between models \cite{Zhang_2025avengers, ong2025routellm, mohammadshahi2024routoo}, enabled features (e.g. retrieval or tool use) \cite{jeong-etal-2024-adaptiverag}, or a combination \cite{su2026toolorchestra}. %
Despite cost and latency improvements, these systems often struggle in distribution shifts \cite{shnitzer2024llmroutingbenchmark, wang2024benchcoe} and reaching substantial improvements over the single-best model \cite{mohammadshahi2024routoo, srivatsa-etal-2024-llmrouting, wang2024benchcoe, wang-etal-2025-mixllm}. Higher oracle performance does not always lead to greater achieved performance \cite{srivatsa-etal-2024-llmrouting}. The candidate lists are occasionally motivated by model size disparities \cite{ong2025routellm} or individual performance \cite{mohammadshahi2024routoo, wang2024benchcoe}), but routers are otherwise often expected to learn complementary (or redundant) model expertise from a broad line-up \cite{mohammadshahi2024routoo, hari2023tryage, wang-etal-2025-mixllm, ICLR2024_frugalfoe}. Of these studies, only two show substantial increases in performance \cite{hari2023tryage, ICLR2024_frugalfoe}.

\paragraph{After-generation MAS} If compute is no concern, one can elicit multiple LLM generations to collapse into one final response. At its simplest, this can be selected via majority voting \cite{li2024moreagents, wang2023selfconsistency} or an LLM judge \cite{toshniwal2025genselect, chen2026peerreviewllm}, but generations can also be fused together \cite{jiang-etal-2023-blender,huang2024deepen}. At greater complexities, cascades of LLM responses can build upon another \cite{virtuallab, su2026toolorchestra, chen2026llmensemblesurvey, ICLR2025_mixtureofagents} or in other structures \cite{masgpt_pmlr2025, liu2024dyLAN}. Most analyses of after-generation MAS look only at homogeneous systems: increasing the number of agents monotonically improves performance, regardless of structure \cite{li2024moreagents}. Such studies find further improvement by varying base model prompts \cite{virtuallab, NEURIPS2023_camelroleplay, masgpt_pmlr2025, pmlr-gptswarm}, enabled features \cite{agentswarm} or hierarchical structure \cite{masgpt_pmlr2025, pmlr-gptswarm, liu2024dyLAN}. However, the benefits of a homogenous MAS (over a single-model) may depend on interactions between the task, base model, and organisation \cite{kim2026sciencescalingagentsystems}. Recent work suggests that increased diversity, via differing prompts or base models, is required to prevent saturation of a MAS \cite{yang2026diversity}, yet too much diversity in a heterogeneous MAS, though good for many tasks, may reduce reasoning performance \cite{abdulaal2025DMoA}. Other heterogeneous MAS have conflicting results; DeePEn does not see monotonic improvement with more agents, and LLM-Blender has unstable performance across datasets \cite{huang2024deepen,jiang-etal-2023-blender}. %

Overall, we see instability in heterogeneous MAS. While there is conflicting evidence on the ideal size of the candidate pool for after-generation MAS, there is no investigation on the ideal models to include in a candidate pool for either before- or after-generation MAS, whereas in classical ensemble systems, this was a central component of ensemble creation.

\section{Method}\label{sec:method}

We focus on science reasoning tasks, which span multiple domains and thus require a diverse set of reasoning skills and abilities. We select three recent, difficult benchmarks to evaluate individual and combined model performance: Humanity's Last Exam (HLE) \cite{hle}, GPQA-Diamond (GPQA) \cite{rein2024gpqa} and Frontier Science--Olympiad (FS) \cite{wang2026frontierscienceevaluatingaisability}. We define our space of all available models $\hat{\mathcal{M}}$ from 23 different LM agents listed in \cref{tab:models}. These models were released between 2024 and 2026, span across 6 different architecture families, and range from 2B to 1.6T parameters, dense and mixture-of-experts and reasoning and non-reasoning models, including 4 science-specialised models. For each question, we obtain 5 generations from each model $m\in\hat{\mathcal{M}}$. As HLE and FS are open-form question-answering datasets, we evaluate response equivalence to the correct answer using gpt-oss-120b as our judge.\footnote{We compare our judge's evaluations to a human annotator on a subset in \Cref{app:judge} and agree on 93\% of decisions (Cohen's kappa: 0.63).}

In cases where training or calibration data is required (to train our router or to determine relative model performance), we construct a training set of 15.5k questions (24.9\% of which are multiple-choice questions). These are relatively equal subsets of AOPS \cite{aopsdataset}, Turing, Scale and Stack-Overflow \footnote{These are closed source STEM reasoning datasets obtained from \href{https://www.turing.com/advance/datasets/stem}{Turing}, \href{https://scale.com/}{Scale} and \href{https://stackoverflow.co/data-licensing/}{Stack-Overflow}}. We evaluate its similarity to our test benchmarks in \Cref{app:train}, and show that relative model performance is highly correlated ($r>0.9,p<10^{-5}$) across all train and test subsets.

\paragraph{Model signals} \textit{Can we approximate MAS and relative model performance from information known from readily available model information, or must we always first evaluate our candidate models?} In classical ensemble models, models were specially trained for response diversity; however, this is not always possible when working with pre-trained models. From most model cards, we can extract 3-4 \textit{pre-evaluation model signals} (1) size, (2) release date, (3) architectural family, and occasionally (4) domain specialization. Using a calibration set, we can evaluate for three \textit{}{post-evaluation model signals}: (1) accuracy, (2) correct-answer diversity, and (3) error diversity.  We approximate correct answer diversity via the Jaccard distance of the pass@1 success set of each model on the training data. Error diversity is measured via the pairwise distance of the pass@1 incorrect responses of each model-- to reduce noise from mismatched clustering, we look only at multiple-choice questions. We convert all measures into normalized pairwise distance matrices\footnote{For model size differences, we calculate differences in log-transformed billion parameters}, and use Mantel tests ($r_M$) and Spearman's correlation ($r_s$) to assess for correlations. Finally, we perform hierarchical agglomerative clustering of the two distance metrics (correct answer diversity and error diversity) with average linkage and look for architectural patterns in the obtained clusters. To evaluate domain specialization, we look at relative performance of our physics and chemistry specialists to comparable generalist models across annotated subsets of our training data. We compare models of shared base model (Llama-3.1-8B, cosmosage-v3.1, and Chem-R-8B), as well as models with comparable performance of differing architectures (OLMo3-7B, cosmosage-v3.1, ChemDFM-R-14B).

\paragraph{Oracle MAS} \textit{How much of an improvement in performance can we expect, given an ideal MAS?} We evaluate the maximum pass@1 score for the entire system ($C_{all} =\hat{\mathcal{M}}$). We also evaluate the performance of differing subsets $C\subset\hat{\mathcal{M}}$, where $|C|=k,k\in\{3,5,10,15,20\}$. We evaluate 8 different methods ($s$) of selecting $C\subset\hat{\mathcal{M}}$. For each selection method $s$, we obtain an ordering of $\hat{\mathcal{M}}$ and define
$C_{s,k}$ as the subset containing the top $k$ models under this ordering. We include both pre- and post-evaluation metrics. We first list pre-evaluation metrics, followed by post-evaluation metrics. \begin{enumerate}[noitemsep, topsep=3pt]
    \item \textbf{Size} \Siz \hspace{6pt} We sort models by their \textit{total parameter size} (which can approximate knowledge capacity and reasoning power \cite{ICLR2025_knowledgecap_scalinglaws,srivastava2023beyond}). To limit interaction of model architecture, we only take one model per family.
    \item \textbf{Family} \Fam \hspace{6pt} We group models by \textit{shared model architectures}. In this instance, we do not take subsets of varying sizes, but instead include all members within the same family. We evaluate: (1) Olmo, (2) Llama, (3) Qwen3, (4) Qwen3.5, (5) gemma and (7) gpt-oss family groups.
    \item \textbf{LLM Chosen} \Llm \hspace{6pt} Given descriptions of all models in $\hat{\mathcal{M}}$ and the dataset, an XLLM (GPT-5 with Deep Research enabled) selects the top $k$ models.
    \item \textbf{Accuracy} \Acc \hspace{6pt} We sort models by their average \textit{accuracy} (pass@1 performance).
    \item \textbf{IoU} \IoU \hspace{6pt} We sort models by their \textit{correct answer diversity}.
    \item \textbf{Error} \Err \hspace{6pt} We sort models by their \textit{error diversity}.
    \item \textbf{Accuracy$\times$Err \AxE/Accuracy$\times$IoU \AxI}  We \textit{optimise} for both accuracy and diversity by applying a 50\% weighting to both values. We start with the most accurate model, and then harmonise the Accuracy and IOU/Error scores.
\end{enumerate}

In the case of Size and Family, we do not evaluate for all possible values of $k$, and only take the sizes of $k$ possible given the outlined restrictions. We also take 5 random subsets of $C_k \subset \hat{\mathcal{M}}$ as baseline and report the overall performance (in Accuracy) and well as the change in performance from the the top-performing model within each $C_{s,k}$ ($\Delta$MAS Gain).

\paragraph{Achieved MAS} \textit{How much of our prophesized MAS is achievable with our architectural set up?} We evaluate our MAS subsets on three different system types, comprising both before- and after-generation systems (See \Cref{fig:mas_types}): \begin{enumerate}[noitemsep, topsep=3pt]
\item \textbf{Routing} is our only before-generation MAS, where we train a simple clustering model, following AvengersPro \cite{Zhang_2025avengers, li2026llmrouterbench}, to optimise for pass@1 accuracy on our training data.
\item \textbf{Majority vote} is the simplest after-generation MAS. We embed all five final-answer generations from each model in $C_{s,k}$ using E5-Large-V2\footnote{We evaluated all-MiniLM-L6-v2, bge-large-en-v1.5 and E5-large-v2 as embedding models and chose the embedding model that gave the highest average accuracy.} and cluster all responses using agglomerative clustering with a maximum cosine distance threshold of 0.15 (We evaluate the impact of this maximum threshold in \Cref{app:cosine_dist}). We then evaluate the correctness of the majority vote via the judgement of the datapoint closest to the cluster's centroid.
\item \textbf{LLM Judge} is a more complex after-generation MAS, where we implement GenSelect \cite{toshniwal2025genselect} using gpt-oss-120b as our final judge. As we have 5 generations per up to 23 models being evaluated, we stratify the judging process due to context-length concerns; the judge evaluates the best answer within the set of generations for each model, and then within subsets of up to 8 models at a time in a tournament-style, until it has selected one singular best response.
\end{enumerate}
We report performance of each MAS in comparison to the single-best model of the system ($\Delta$MAS Gain) in the main text to highlight the relative benefits/downsides of using a MAS over one model already in the system. However, we report overall accuracies in \Cref{app:actual}. %
The single-best performance is calculated separately for each $C_{s,k}$, depending on the contained models, and MAS, to allow for comparability: For routing, we report the pass@1 of the best model. For majority vote, we report the best majority@5. For LLM judge, we ask our judge to pick the best generation per model, and report the best average accuracy.

\section{Results}

In the main article, we visualise results for HLE, as it is the largest and most challenging dataset; however, performance is similar across all datasets, and we report important differences in the main text. You can see all results in \Cref{app:oracle} and \Cref{app:actual}.

\subsection{Model signals}
\begin{figure*}[ht]
    \centering
\includegraphics[width=\linewidth]{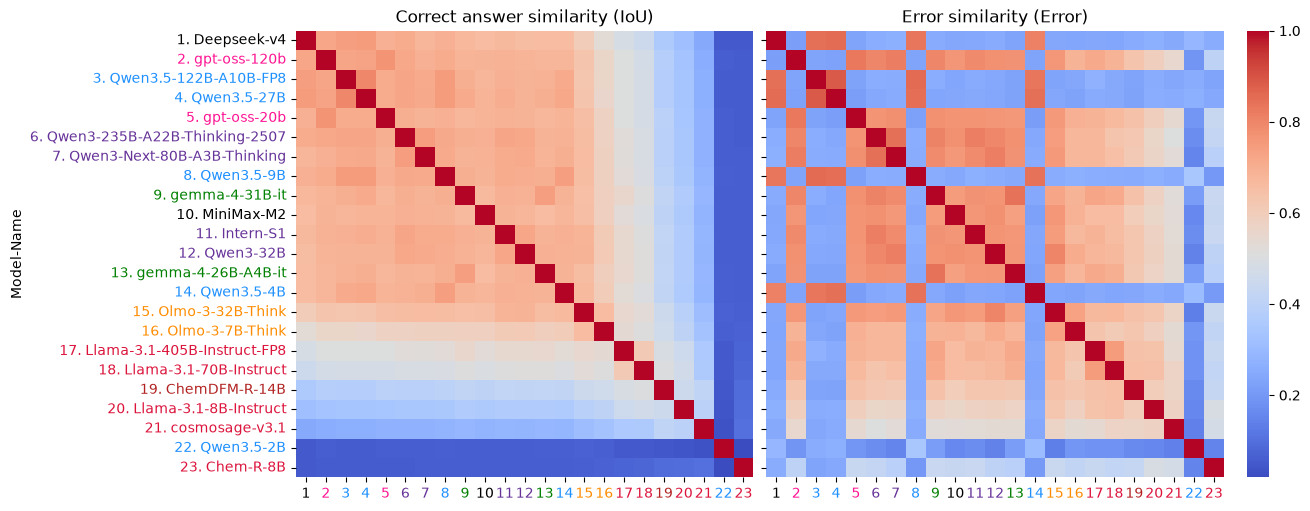}
    \caption{Models are sorted in order of decreasing accuracy on the training data (the text is coloured by architectural family). The colour of each cell indicates the pairwise similarity of model responses (left being correct answers, right being incorrect answers). More accurate models get similar questions correct ($r_M=0.931,p<.01$), yet their errors are only mildly correlated ($r_M=0.384, p<.01)$.}
    \label{fig:Diversities}
\end{figure*}

We present individual model behaviours on our training data in \Cref{fig:Diversities}. Accurate models tend to get similar questions correct ($r_M=0.931,p<.01$). However, their errors are only weakly correlated ($r_M=0.384, p<.01)$. Model size has a moderate correlation with accuracy ($r_s=0.583,p<.01$) and weak correlation with correct answer similarity ($r_M=0.219, p<.05)$. When performing agglomerative clustering (See \Cref{app:extra} for visualizations), we can see some patterns emerge: Models trained with reasoning have greater similarity of correct answers (three exceptions are Qwen3.5-2B, Chem-R-8B, and ChemDFM-R-14B; the latter two have very different post-training in comparison to the other reasoning models). When clustering by error similarity, we see architectural patterns: Deepseek-v4 and Qwen3.5 models make very similar errors which are distinct to the errors made by gemma, gpt-oss, OLMo3, Qwen3 and MiniMax.

\begin{figure}[h]
    \centering
\includegraphics[width=\linewidth]{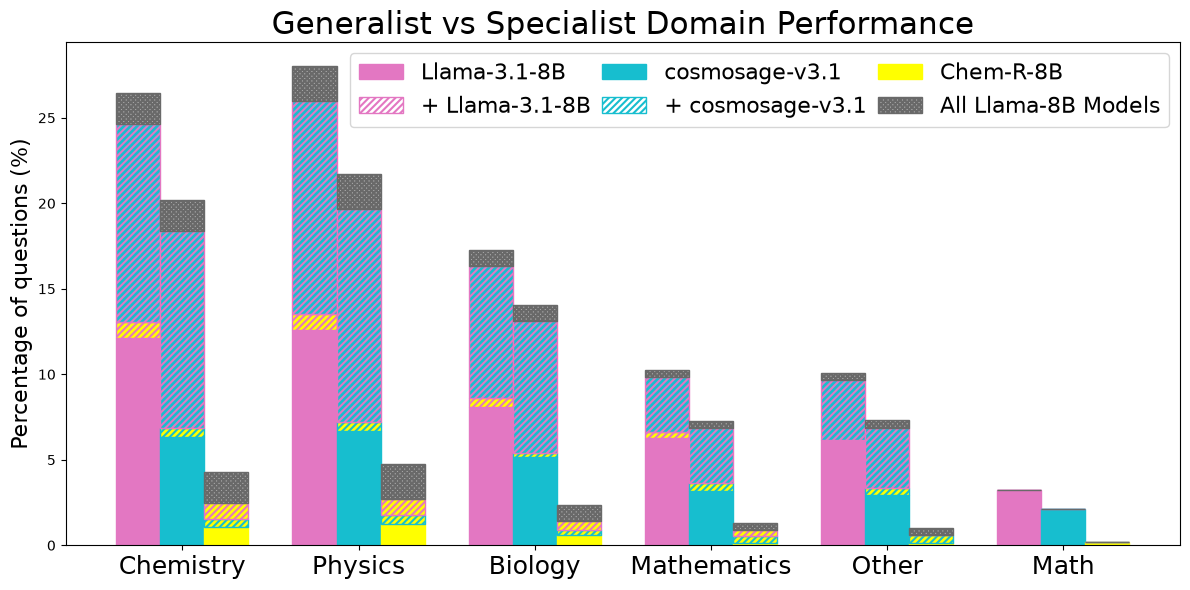}
    \caption{Llama3.1-8B shows just as strong, if not greater performance in all domains in comparison to its Physics and Chemistry-specialised counterparts, especially on Physics and Chemistry questions.}
    \label{fig:GvS}
\end{figure}

We compare performance of generalist versus specialist models in \Cref{fig:GvS,app:GvS}; compared to its physics and chemistry-specialised counterparts (all three models are fine-tuned from Llama-3.1-8B-Base), the normal Llama performs better on all science reasoning domains, particularly in the specialised domains of physics and chemistry. The gains of each specialised model (i.e. questions answered only by the specialist model and not the generalist base), are not limited to the specific specialisation of each model, but are relatively equally divided across all domains. We see a similar pattern when architectures are unrelated (\Cref{app:GvS}). This means that one cannot assume high correct answer diversity (or specialisation) from training data dominance.

Through these model investigations, we find larger reasoning models typically perform better, and have lower correct answer (IoU) diversity. Error diversity seems to vary much more between architectures, though some model families behave more similarly than others. IoU diversity cannot be assumed simply from training data specialisation-- models specially fine-tuned in specific domains do not see greater performance on these domains on our science reasoning subset (regardless if they arise from the same architectural backbone or not). While we see a correlation between size and accuracy, we continue with both forms of subset generation to evaluate for equivalent MAS performance. We list all subset identities in \Cref{app:subsets}.
\subsection{Oracle MAS performance}

\begin{figure*}[ht]
    \centering
    \includegraphics[width=\linewidth]{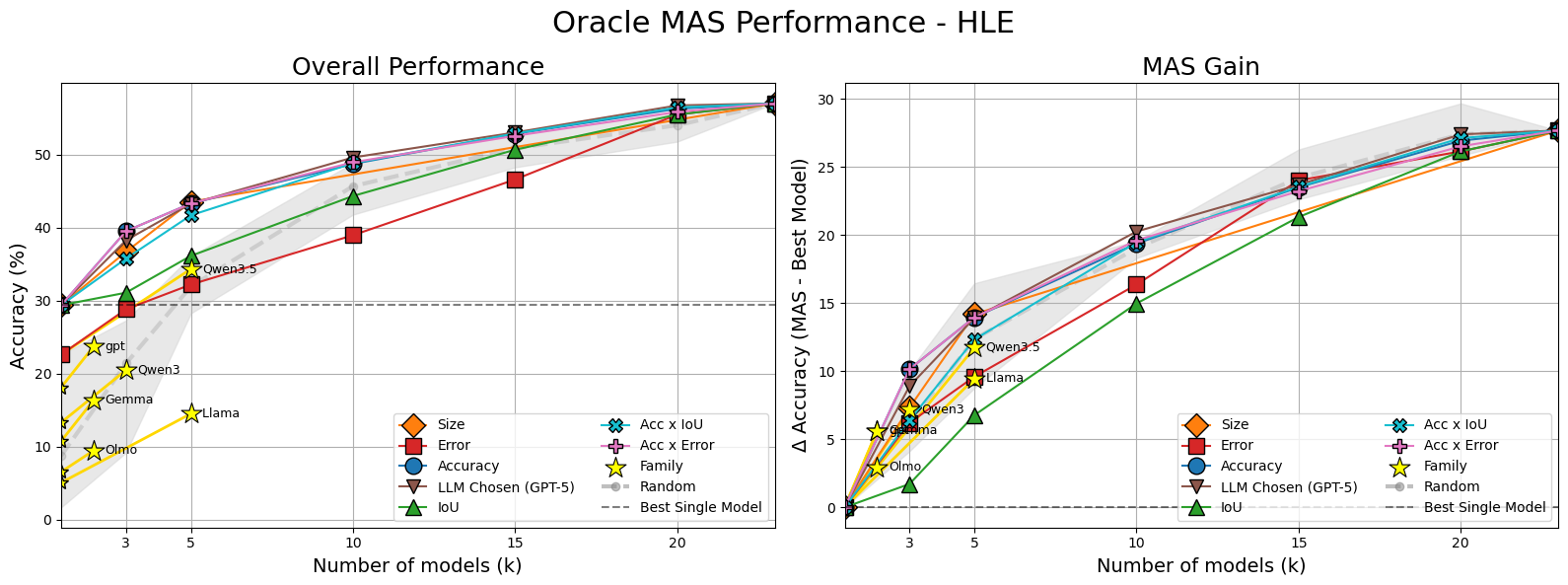}
    \caption{Potential gains of each MAS with increasing $k$. More models consistently indicates greater MAS performance; these gains are strongest when we combine the most accurate or largest models.}
    \label{fig:oracle_mas}
\end{figure*}

In \Cref{fig:oracle_mas} we show the optimal performance of each MAS.\footnote{To showcase random variance, we report the mean performance across 5 runs of Random model groupings, and shade in the range of performance observed in these 5 runs.} Naturally, we observe an increase in performance with increasing $k$. When looking solely at overall accuracy, we see the \textit{greatest potential performance} across all $k$ when grouping models by accuracy (or accuracy combined with some form of diversity) \Acc \AxE \AxI \hspace{3pt} or by LLM suggestion \Llm.
We see the \textit{smallest prophesized improvements} with Error \Err \hspace{3pt} or IoU \IoU \hspace{3pt} diversity; this pattern is conserved across datasets. Notably, though family \Fam\hspace{3pt} grouping shows relatively low overall performance, we can see that some family groups prophesise high MAS gains over a single base model (Particularly gemma models). 

\subsection{Achieved MAS Performance}

\begin{figure*}[h]
    \centering
    \includegraphics[width=\linewidth]{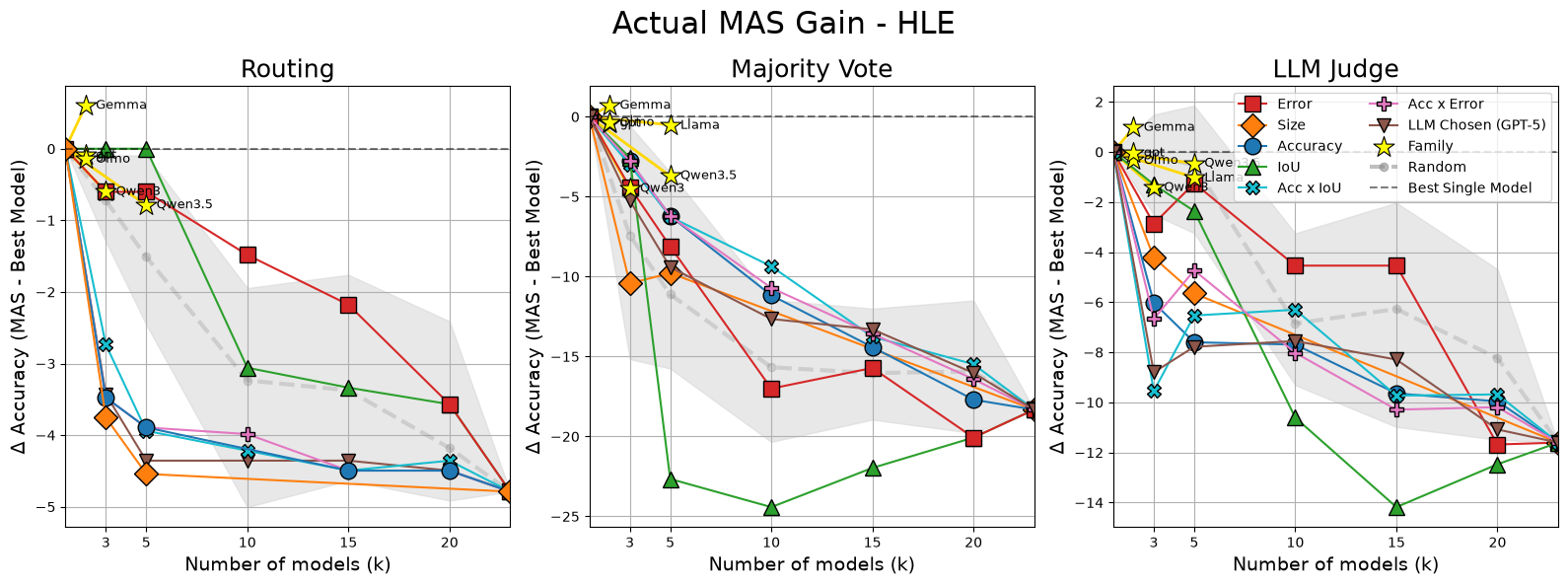}
    \caption{Actual gains of each MAS subset and architecture with increasing $k$. With larger candidate pools, we see larger losses in performance (compared to the single best model of the set). On HLE, Only MAS with models from one model family \Fam\hspace{3pt} show improvements over the baseline best model.}
    \label{fig:actual_mas}
\end{figure*}

For actual MAS performance in \Cref{fig:actual_mas}, one pattern is clear: more models nearly always decreases performance.  Most intentional groupings give performance lower than the baseline best model, and often lower than a random set of models. Across all MAS forms, families \Fam\hspace{3pt} with shared architectures perform best; in most cases, this means the least detriment to performance rather than a large improvement. There is also variability between families. We see typically the best relative MAS performance with Gemma4 MAS, though highly variable performance with OLMo3 and Qwen3.

Routing systems seem to benefit most with greater diversity of correct answers (IoU \IoU\hspace{3pt} has the greatest performance, though it prophesized relatively low improvement with the oracle)-- this behaviour is inline with the training approach of classical routed systems. Only IoU and Family architectures showed a positive improvement on routed structures.

After-generation approaches (Majority-vote and LLM-as-a-Judge) have steep declines in performance with increasing $k$ in comparison to the routed system. It seems intuitive that, with increasing response diversity, the majority-vote accuracy would decrease, given the lowered robustness of the system. Despite an increase in the best single-agent systems using majority vote (29.4\% $pass@1_{Best}(m)$ to 32.2\% $majority@5_{Best}(m)$  on HLE), nearly all heterogeneous MAS groups decline in performance; this is strongest with Error \Err \hspace{3pt} and IoU \IoU \hspace{3pt} diversity in FS and GPQA in \Cref{app:actual}. The only grouping approach that occasionally improves over the single-agent system is within a model family (Gemma4 and OLMo3). However, in some instances (See \Cref{app:cosine_dist}), we see MAS Gain for Accuracy (\Acc) and Error (\Err)-related groupings (\AxI \AxE).

Interestingly, using an LLM-as-a-Judge does not improve MAS stability much (Though we see further increase with the homogenous system, $judge@5_{Best}(m)=36.5\%$ on HLE) Nearly all of our intentional subsets perform worse than random baseline with a LLM-as-a-Judge architectures. This is only seen for HLE-- in GPQA and FS, we see performance approximately equivalent to random baseline. In addition to improved performance on HLE with Gemma4 family architeture, we find a small MAS gain (+$2\%$ on FS, +$<1\%$ on GPQA) with Accuracy \Acc\hspace{3pt}and Accuracy-optimised metrics (\AxI\hspace{3pt}\AxE)

\section{Discussion}
\paragraph{There can be \textit{too much} diversity} Increased diversity may not be best for all MAS architectures, or they may require diversity in different forms. High \textit{solution diversity} pollutes the answer space and may make it difficult to isolate the correct response, especially in reasoning tasks \cite{abdulaal2025DMoA}. Other works finding success with increased diversity looked at a limited amount of diversity over a large amount of agents (e.g. system prompts \cite{virtuallab}, enabled features \cite{agentswarm} or a limited number of base models (three) \cite{yang2026diversity}). Rather than high solution diversity, heterogeneous MAS may require \textit{other forms of diversity}, such as high reasoning diversity but consistent output \cite{wang2023selfconsistency}--which can arise from \textit{diverse prompting or tools, rather than many models}. This ensures some diversity as well as equivalent performance between models \cite{huang2024deepen}. Other studies found that decentralised (i.e. voting) approaches with mixed-capability models showed the greatest improvements \cite{kim2026sciencescalingagentsystems}-- however this behaviour was only measured \textit{within model families} and may improve with more model interaction, which was outside of our study's scope. Similarly, a stronger central orchestrator (i.e. judge) may recover failures in heterogeneous systems, and may need more complex architectural systems to accommodate increased diversity from additional base models.

\paragraph{Routing systems may require more than just correct-answer diversity} Mixture-of-Experts systems, which are routing systems, require correct-answer diversity between their experts \cite{moe_ensemble}. Appropriately, we see highest performance with our IoU-diverse router-- but this is rarely sufficient to increase performance over the base model. For a clear training signal, this per-model expertise must capture different \textit{tasks} \cite{moe_ensemble}, which may not necessarily be measured with our naive IoU distance. Highly successful routing approaches like FrugalFoE,  rely on separately fine-tuned models or models with relatively stronger performance on annotated benchmarks; relying on models determined from annotated benchmarks performed worse than models expressly fine-tuned with domain specialisation \cite{ICLR2024_frugalfoe}. This creates two issues: (1) annotating benchmarks, and (2) actually finding models with the intended specialisations in the haystack. In our work, we found that models trained on different domain subsets do not provide effectively distinct expertise on the different domains. Furthermore, too much overlap between model expertise also pollutes the training signal \cite{NEURIPS2025_MoEoverlap}-- therefore, combining strong, generalist models will not necessarily lead to performance above baseline. Other approaches that successfully identified model expertise ensured all candidate models share a similar size \cite{Zhang_2025avengers} or model architecture \cite{hari2023tryage}. For actual performance gains, there may be additional criteria to include besides only high correct-answer diversity. While we tried to jointly optimise for \textit{high} accuracy, other criteria can be investigated, such as \textit{similar} accuracy, size or architecture. These criteria may restrict one's pool of models, especially with growing generalist models: there may be fewer true specialist models allowing easy integration into a Routed-MAS. Models may instead benefit from specific training or fine-tuning for integration.

\section{Conclusion}

In this work, we evaluate the contribution of model candidate selection on MAS performance. We compare three simple forms of MAS: Routing, Majority vote and LLM-as-a-Judge on three complex science reasoning benchmarks.
While response diversity was explicitly trained for in classical ensemble systems, we must now evaluate our candidate models to determine the ideal pool, which has not been appropriate investigated in previous work. We compare architectural information (size, family) as well as evaluated metrics (accuracy, correct answer diversity, error diversity) as methods to determine model subset lists. We find increasing candidate pool size always seems to impair MAS performance (in contrast to previous work \cite{yang2026diversity}). Groupings with limited prophesized improvements (as measured via oracle performance, such as shared model families or correct answer diversity), often show the best actual performance across MAS architecture. While we see limited improvement in performance from our sample of investigations, we do not believe that means heterogeneous MAS is futile; other studies have reported improvements in performance, even using the same MAS architectures \cite{Zhang_2025avengers}. Though we saw a decrease in performance with increasing architectural complexity for our hetereogenous MAS systems, our baseline homogeneous MAS systems performance \textit{increased} with MAS complexity, suggesting that these architectures may be best suited for homogeneous systems, necessitating the development of architectures specifically for heterogeneous MAS. We strongly recommend MAS engineers to evaluate the contribution of their \textit{choice of models} to their system's performance-- specific criteria may be needed for success of their MAS architecture, and observed failures may arise simply due to a poor choice of candidate models.

\section*{Limitations}
We focus only on before and after-generation multi-agent systems. While during-generation systems also exist, they are out of scope for this paper, and would require a combination of aspects from both before- and after-generation MAS. While there are many other possible implementations of before- and after-generation MAS, we hope that our coverage of three forms provides some insight into MAS-level instabilities in model selections, and \textit{future work can explore the contribution of model selection on other, more advanced MAS architectures.}

We intentionally chose simple MAS architectures. However, each come with their own limitations. We selected only one routing architecture, out of many possible backbones (we chose a recent model that showed high perofmance). Furthermore, while we do evaluate the relative impact of some of our majority-vote design choices in \Cref{app:cosine_dist}, there are multiple ways to evaluate majority-vote, such as semantic clustering using NLI models. Given the complexity of our benchmarks, we chose to embed each response using a large model trained on scientific text, though we did not specifically train an embedding model for this task. Context-length issues may also contribute to the poor observed performance of our LLM-Judge MAS, GenSelect, on high values of $k$. This may be improved with more advanced stratification of the voting system. More complicated MAS structures may be able to compensate for the instability introduced with adding more models. However, \textit{we leave improvements of these simple MAS architectures to future work.}

To ensure equal comparison of all models, we do not enable tool use or retrieval, even with models that have that capability. We note that models trained to innately use tools are often trained with different sandbox environments, which can differentially advantage or disadvantage different models. However, tool use is especially helpful for many questions within the benchmarks we investigate, and is very common in MAS deployment. Therefore, \textit{future work can explore how tool use can impact response diversity and heterogeneous MAS performance}

Given that we are limited by the number of available models within each model family, our Family \Fam\hspace{3pt} investigations cannot test for all levels of $k$. Future work, with more access to larger model families, or other forms of architectural grouping, \textit{could see if larger family-based groupings could increase performance with greater k.}

While we do look at a large sample of models, this could be expanded even larger. However, we observe limited gains with the number of models we included, therefore, we did not further increase the candidate set size. We tried to diversify our model pool across model family, size, year of release, and training architecture, but \textit{further work could evaluate more model families and expand our investigations by also comparing between reasoning training types, prompts and enabled features.}

We look specifically at scientific reasoning tasks evaluated by AAI-- there are other difficult reasoning tasks, like mathematics and coding. However, single model performance is much lower on scientific reasoning tasks (like physics and chemistry) than these domains, giving a greater overhead for MAS over single-agents, motivating our choice. However, \textit{future work can see how these results compare across diverse reasoning domains.}

We use one judge for all of our experiments when needed (gpt-oss-120b). We maintain this consistent judge for simplicity, and use our second-best performing model (Deepseek-v4-Pro was only released in late April 2026. The majority of this work was completed beforehand.). We do validate the quality of our judge in \Cref{app:judge}, and note that the judge is more likely to judge a question as correct than our human grader. However, the choice of judge could impact relative model performance and LLM-as-a-Judge MAS performance. \textit{Future work can investigate how choice of judge in a heterogeneous LLM-as-a-Judge system can impact MAS performance.}

\bibliography{main}

\appendix
\section{Extra analyses}
\label{app:extra}

\subsection{All models}\label{app:models}

We present all models $m\in\hat{\mathcal{M}}$ in \Cref{tab:models}. 
\begin{table*}[ht]
\centering
\scriptsize
\setlength{\tabcolsep}{2pt}
\renewcommand{\arraystretch}{0.95}
\begin{tabularx}{\textwidth}{XrrX}
\toprule
\textbf{Model} & \textbf{Params(B)} & \textbf{Year} & \textbf{Notes} \\
\midrule
Deepseek-v4-Pro \cite{deepseekv4} & 1600/49 & 2026 & \thinking\moe \\
Qwen3.5-122B-A10B-FP8 \cite{qwen3.5} & 122/10 & 2026 & \thinking \\
Qwen3.5-27B \cite{qwen3.5} & 27 & 2026 & \thinking \\
Qwen3.5-9B \cite{qwen3.5} & 9 & 2026 & \thinking \\
Qwen3.5-4B \cite{qwen3.5} & 4 & 2026 & \thinking \\
Qwen3.5-2B \cite{qwen3.5} & 2 & 2026 & \thinking \\
gemma-4-31B-it \cite{gemma4} & 31 & 2026 & \thinking \\
gemma-4-26B-A4B-it \cite{gemma4} & 26/4 & 2026 & \thinking\moe \\
MiniMax-M2 \cite{minimaxm2} & 230/10 & 2026 & \thinking\moe \\
Olmo-3-32B-Think \cite{olmo3} & 32 & 2026 & \thinking \\
Olmo-3-7B-Think \cite{olmo3} & 7 & 2026 & \thinking \\
ChemDFM-R-14B (\qwen) \cite{chemdfm} & 14 & 2026 & \thinking\science\\
gpt-oss-120b \cite{gptoss} & 120/5.1 & 2025 & \thinking\moe \\
gpt-oss-20b \cite{gptoss} & 20/3.6 & 2025 & \thinking\moe \\
Qwen3-235B-a22B-Thinking-2507 \cite{qwen3} & 235/22 & 2025 & \thinking\moe \\
Qwen3-Next-80B-A3B-Thinking \cite{qwen3} & 80/3 & 2025 & \thinking\moe \\
Qwen3-32B \cite{qwen3} & 32 & 2025 & \thinking \\
Intern-S1 (\qwen) \cite{interns1} & 235/22 & 2025 & \thinking\moe\science \\
Chem-R-8B (\llama) \cite{chemr} & 8 & 2025 & \thinking\science \\
cosmosage-v3.1 (\llama) \cite{cosmosage} & 8 & 2024 & \science \\
Llama-3.1-405B-Instruct-FP8 \cite{llama3} & 405 & 2024 & \\
Llama-3.1-70B-Instruct \cite{llama3} & 70 & 2024 & \\
Llama-3.1-8B-Instruct \cite{llama3} & 8 & 2024 & \\
\bottomrule
\end{tabularx}
\caption{Overview of language models evaluated. \thinking\ indicates the model is run with reasoning mode on (if available), \science\ indicates the model is additionally trained on science-specific data, \moe\ indicates the model is a Mixture-of-Experts model. For some additionally fine-tuned models, we note the model backbones: \qwen\ indicates a Qwen-backbone, whereas \llama\ indicates a Llama-backbone.}
\label{tab:models}
\end{table*}

\subsection{Judge verification}\label{app:judge}
To evaluate the quality of our LLM Judge (gpt-oss-120b), we take a random subsample of 100 answers for HLE (for any model in $\hat{\mathcal{M}}$) and manually grade them in reference to the provided solution. We obtain an agreement of 93\% and a Cohen's kappa of 0.63, which indicates substantial alignment. Notably, the LLM judge is much more likely to grade an answer as correct-- there is only one instance of the LLM judge grading a human-verified correct answer as incorrect. Therefore, if anything, we report higher accuracies than would be observed in actuality.

\subsection{Majority vote Robustness}\label{app:cosine_dist}
\begin{figure*}
    \centering
\includegraphics[width=\linewidth]{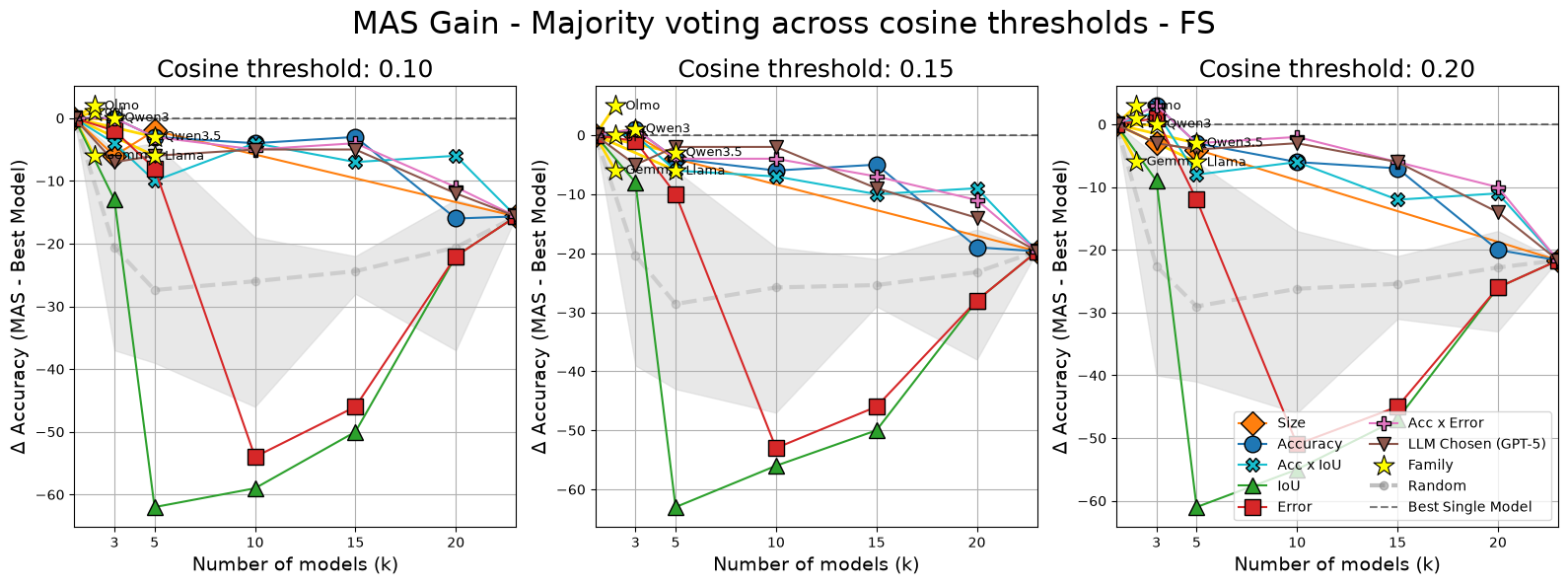}
    \caption{We show the \textbf{MAS gain} across groupings of $C_{s,k}$ with three different cosine distance thresholds on FrontierScience-Olympiad}
    \label{fig:cosine_gain}
\includegraphics[width=\linewidth]{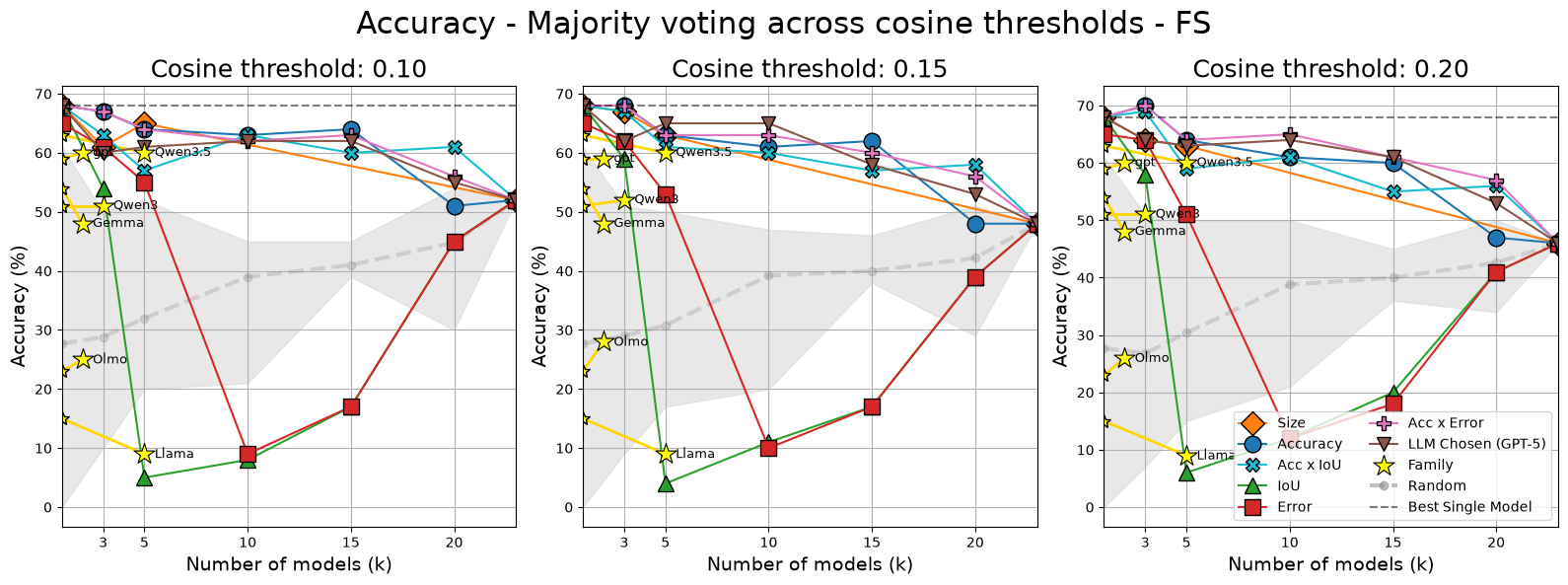}
    \caption{We show the \textbf{overall accuracy} across groupings of $C_{s,k}$ with three different cosine distance thresholds on FrontierScience-Olympiad}
    \label{fig:cosine_acc}
\end{figure*}

The choice of embedding model and cosine distance threshold in implementing majority vote can impact the performance of the MAS system. As noted in \Cref{sec:method}, we evaluated the performance of three embedding models on our datasets: E5-Large-V2, all-MiniLM-L6-v2 and bge-large-en-v1.5. We chose the embedding model that gave the highest average accuracy, across all models. We also assess the relative impact of the choice of cosine distance threshold in \Cref{fig:cosine_gain,fig:cosine_acc}. While we do small differences in accuracies and MAS gain upon varying our choice of a cosine distance threshold (between 0.10, 0.15, and 0.20), the patterns remain largely similar patterns across all selected values. However, we do see small MAS gains for Accuracy \Acc\hspace{3pt},  Acc$\times$Error \AxE\hspace{3pt}, Acc$\times$IoU \AxI\hspace{3pt}, and Error \Err\hspace{3pt}, which suggests more relaxed clustering thresholds could improve performance of this group. However, in the main text, we report with a cosine distance threshold of 0.15, to ensure high semantic similarity of the clustered answers.

\subsection{Training data evaluation}\label{app:train}

\begin{figure}
    \centering
    \includegraphics[width=\linewidth]{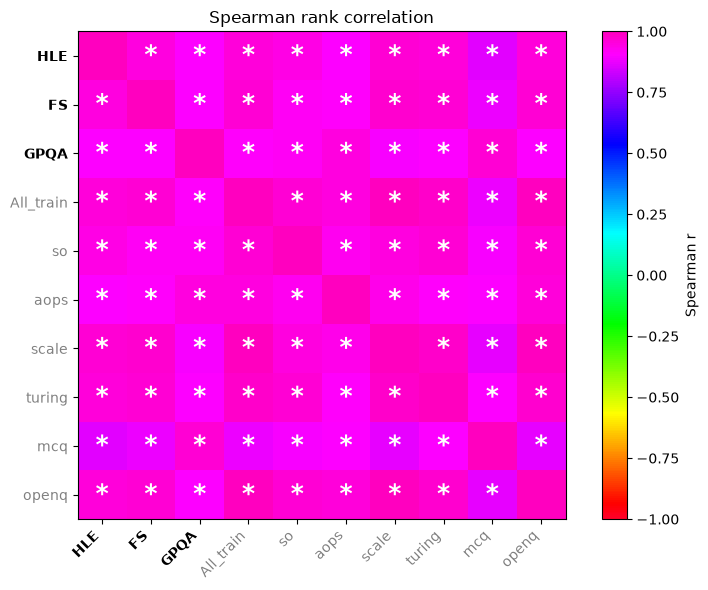}
    \caption{We see high ($p<10^{-5}, r>0.9$) correlation between our training data and all three test datas.}
    \label{fig:data_corr}
\end{figure}

We evaluate the appropriateness of our training data (and its subsets) to our test data to ensure its validity. We test the Spearman rank correlation across all evaluated models and show the results in \Cref{fig:data_corr}. All combinations show high correlation ($p<10^{-5};r>0.9$). 

\subsection{Correct and Incorrect Answer Clustering}\label{app:single}

\begin{figure*}
    \centering
    \includegraphics[width=\linewidth]{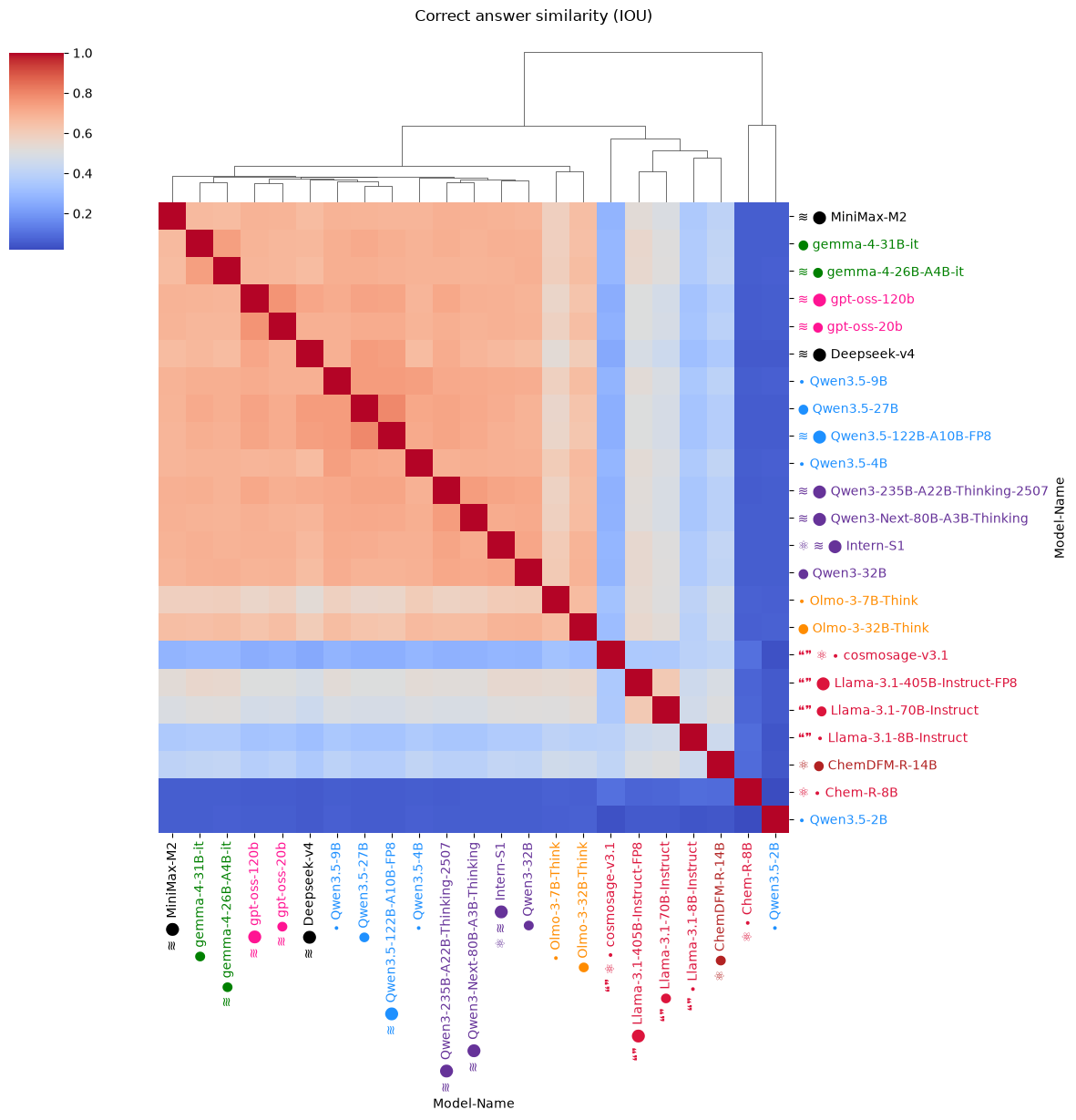}
    \caption{Clustered IoU Performance}
    \label{fig:IoU}
\end{figure*}

\begin{figure*}
    \centering
    \includegraphics[width=\linewidth]{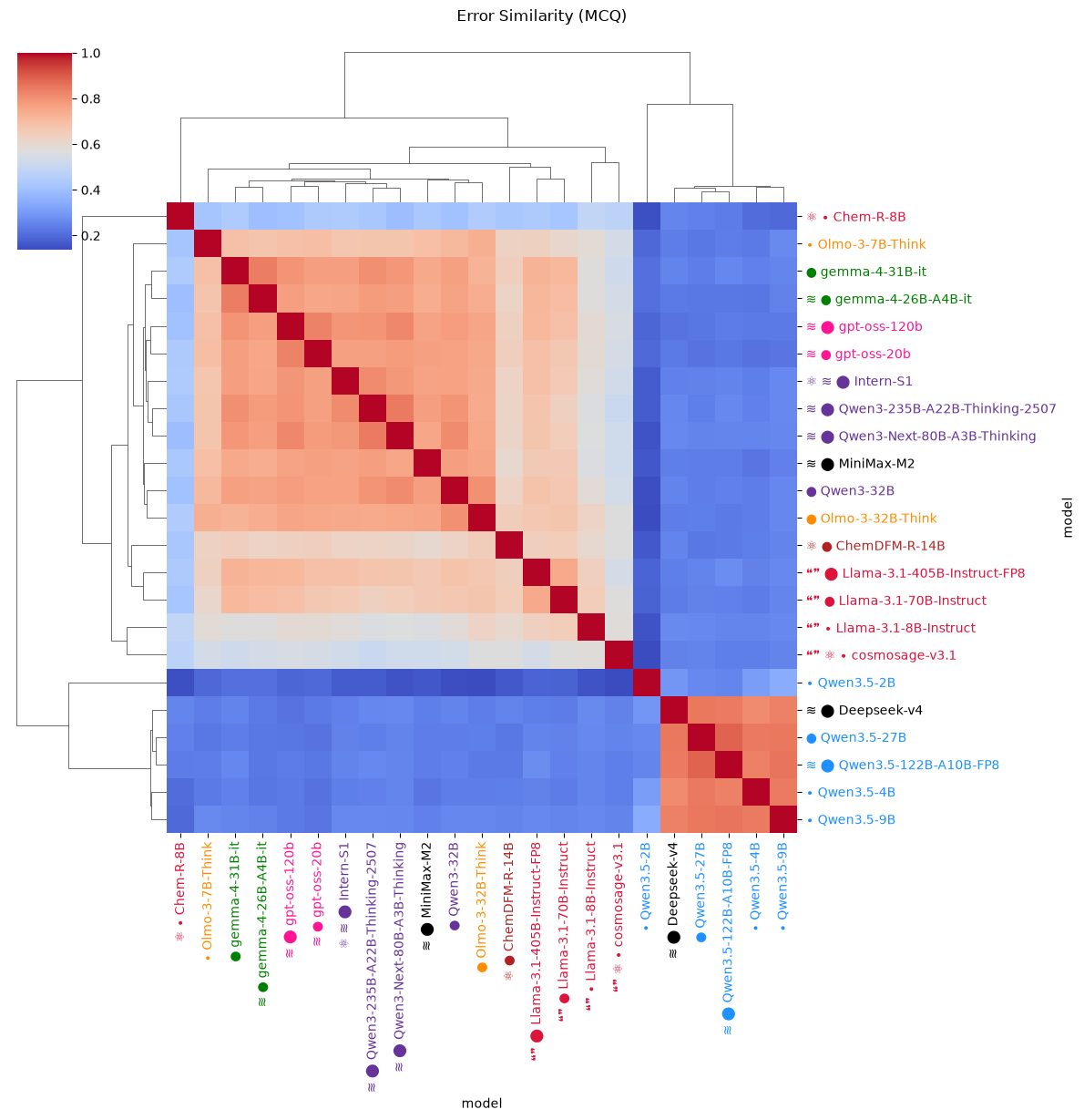}
    \caption{Clustered Error Performance}
    \label{fig:Error}
\end{figure*}

We show the heatmaps, ordered according to the obtained dendrogram from hierarchical agglomerative clustering for correct answer (IoU) similarity in \Cref{fig:IoU} and for error diversity in \Cref{fig:Error}.

In \Cref{fig:IoU}, we see a fairly similar heatmap as to when it was organised by accuracy in \Cref{fig:data_corr}. There is only one, large cluster, with relatively high similarity. This cluster contains most reasoning models (save for OLMo3-7B models, which performs fairly similarly to this cluster). Just outside of this cluster, we see instruction-tuned models. Chem-R-8B and Qwen3.5-2B have very dissimilar performance to all models.

In \Cref{fig:Error}, we see a strikingly different heatmap in comparison to \Cref{fig:data_corr}. There are two clusters. One contains Deepseek-v4 and most Qwen3.5 models, which have very high error similarity ($>0.8$). The second cluster contains most other reasoning model families. Instruction-tuned models are more similar to this second group of reasoning models. Qwen3.5-2B and Chem-R-8B again have very dissimilar performance to all models.

\subsection{Generalist vs Specialist Model Performance}\label{app:GvS}

\begin{figure}
    \centering
    \includegraphics[width=\linewidth]{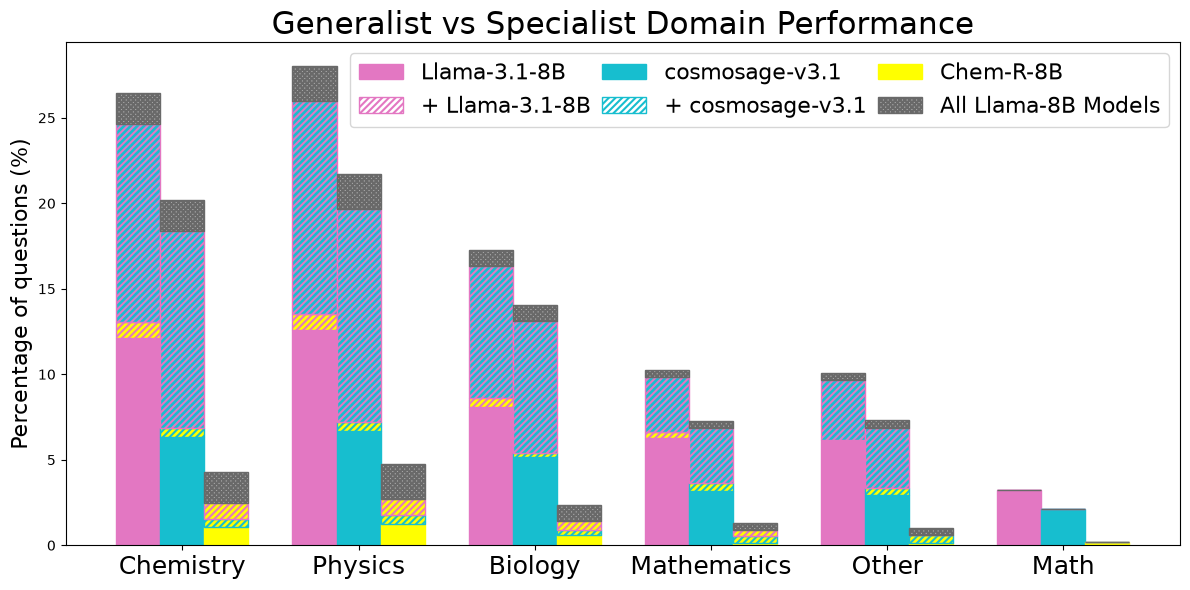}
    \caption{Llama3.1-8B performance is compared to its physics (cosmosage-v3.1) and chemistry (Chem-R-8B) specialised counterparts. All three models originate from the same pretrained model (Llama-3.1-8B-Base).}
    \label{fig:gvs1}
\end{figure}

\begin{figure}
    \centering
    \includegraphics[width=\linewidth]{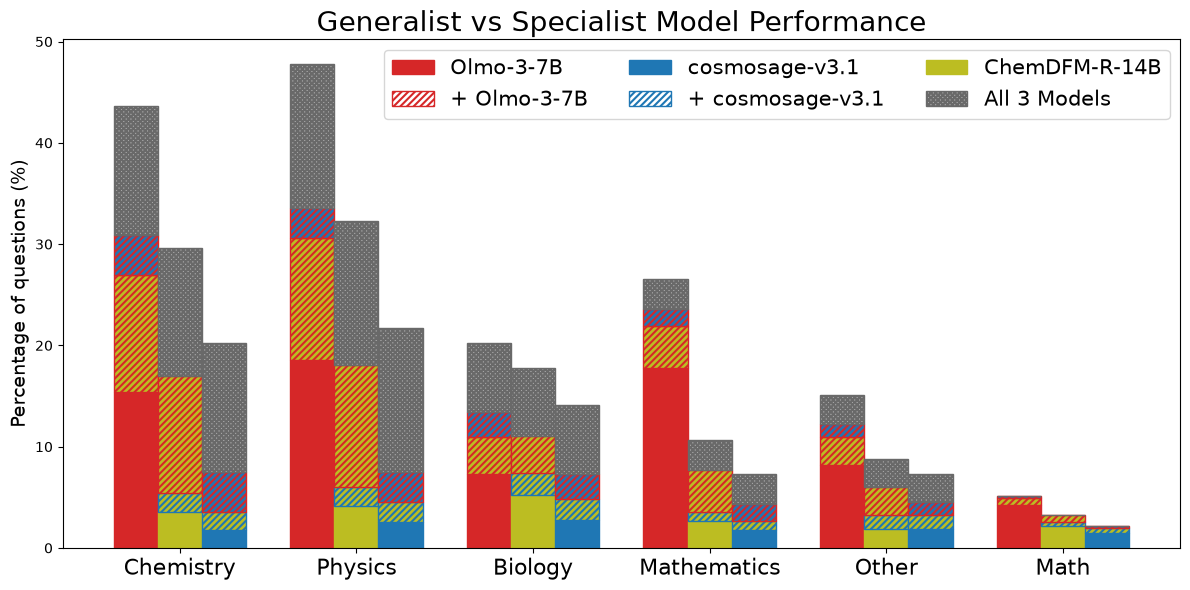}
    \caption{We compare three unrelated generalist (OLMo3-7B-Think) and specialist (ChemDFM-R-14B, cosmosage-v3.1) models. Again we see the generalist model typically outperforms the specialists in every domain, and the specialists do not show a strong preference for their specialised domain's questions. The strongest observed specialisation is the generalist model for Mathematics.}
    \label{fig:gvs2}
\end{figure}

We present the results of comparing generalist and specialist models of a shared architecture in \Cref{fig:gvs1} (as is shown in the main article) and generalist and specialist models of different architectures (but comparable performance) in \Cref{fig:gvs2}. We see that the specialist models typically do not outperform the `generalist' models in their specialised domains, but tend to have relatively equivalent performance across all annotated topics. The most striking difference in performance across topics is our reasoning generalist, Olmo3-7B's, increased performance on mathematics data.

\subsection{Subset Identities}\label{app:subsets}

We present the identities of each model evaluated in each described subset, in each of the evaluated sizes, in \Cref{tab:all_subsets1,tab:all_subsets2,tab:all_subsets3}. All subsets, at least in smaller sizes ($k<10$), have distinct model combinations.

\rowcolors{2}{white}{gray!15}
\begin{table*}
\centering
\scriptsize
\begin{tabularx}{\textwidth}{lX}
\toprule
\textbf{Run-Name}  & \textbf{Model-List} \\
\midrule
All & 'ChemDFM-R-14B', 'gpt-oss-120b', 'Qwen3-235B-A22B-Thinking-2507', 'MiniMax-M2', 'Qwen3-Next-80B-A3B-Thinking', 'Chem-R-8B', 'gpt-oss-20b', 'cosmosage-v3.1', 'Intern-S1', 'Llama-3.1-8B-Instruct', 'Llama-3.1-70B-Instruct', 'Llama-3.1-405B-Instruct-FP8', 'Qwen3.5-122B-A10B-FP8', 'Qwen3.5-27B', 'Qwen3.5-9B', 'Qwen3.5-4B', 'Qwen3.5-2B', 'Olmo-3-32B-Think', 'Olmo-3-7B-Think', 'gemma-4-26B-A4B-it', 'gemma-4-31B-it', 'Deepseek-v4' \\
Accuracy x IoU-3 & 'Deepseek-v4', 'Qwen3.5-2B', 'gpt-oss-120b' \\
Accuracy x IoU-5 & 'Deepseek-v4', 'Qwen3.5-2B', 'gpt-oss-120b', 'Qwen3.5-122B-A10B-FP8', 'Qwen3-235B-A22B-Thinking-2507' \\
Accuracy x IoU-10 & 'Deepseek-v4', 'Qwen3.5-2B', 'gpt-oss-120b', 'Qwen3.5-122B-A10B-FP8', 'Qwen3-235B-A22B-Thinking-2507', 'Chem-R-8B', 'Qwen3.5-27B', 'gpt-oss-20b', 'MiniMax-M2', 'gemma-4-31B-it' \\
Accuracy x IoU-15 & 'Deepseek-v4', 'Qwen3.5-2B', 'gpt-oss-120b', 'Qwen3.5-122B-A10B-FP8', 'Qwen3-235B-A22B-Thinking-2507', 'Chem-R-8B', 'Qwen3.5-27B', 'gpt-oss-20b', 'MiniMax-M2', 'gemma-4-31B-it', 'cosmosage-v3.1', 'Qwen3-Next-80B-A3B-Thinking', 'Qwen3.5-9B', 'Intern-S1', 'gemma-4-26B-A4B-it' \\
Accuracy x IoU-20 & 'Deepseek-v4', 'Qwen3.5-2B', 'gpt-oss-120b', 'Qwen3.5-122B-A10B-FP8', 'Qwen3-235B-A22B-Thinking-2507', 'Chem-R-8B', 'Qwen3.5-27B', 'gpt-oss-20b', 'MiniMax-M2', 'gemma-4-31B-it', 'cosmosage-v3.1', 'Qwen3-Next-80B-A3B-Thinking', 'Qwen3.5-9B', 'Intern-S1', 'gemma-4-26B-A4B-it', 'Llama-3.1-8B-Instruct', 'Qwen3-32B', 'Qwen3.5-4B', 'ChemDFM-R-14B', 'Olmo-3-32B-Think' \\
Accuracy-3 & 'Deepseek-v4', 'gpt-oss-120b', 'Qwen3.5-122B-A10B-FP8' \\
Accuracy-5 & 'Deepseek-v4', 'gpt-oss-120b', 'Qwen3.5-122B-A10B-FP8', 'Qwen3.5-27B', 'Qwen3-235B-A22B-Thinking-2507' \\
Accuracy-10 & 'Deepseek-v4', 'gpt-oss-120b', 'Qwen3.5-122B-A10B-FP8', 'Qwen3.5-27B', 'Qwen3-235B-A22B-Thinking-2507', 'gpt-oss-20b', 'Qwen3-Next-80B-A3B-Thinking', 'Qwen3.5-9B', 'gemma-4-31B-it', 'Qwen3-32B' \\
Accuracy-15 & 'Deepseek-v4', 'gpt-oss-120b', 'Qwen3.5-122B-A10B-FP8', 'Qwen3.5-27B', 'Qwen3-235B-A22B-Thinking-2507', 'gpt-oss-20b', 'Qwen3-Next-80B-A3B-Thinking', 'Qwen3.5-9B', 'gemma-4-31B-it', 'Qwen3-32B', 'Intern-S1', 'MiniMax-M2', 'gemma-4-26B-A4B-it', 'Qwen3.5-4B', 'Olmo-3-32B-Think' \\
Accuracy-20 & 'Deepseek-v4', 'gpt-oss-120b', 'Qwen3.5-122B-A10B-FP8', 'Qwen3.5-27B', 'Qwen3-235B-A22B-Thinking-2507', 'gpt-oss-20b', 'Qwen3-Next-80B-A3B-Thinking', 'Qwen3.5-9B', 'gemma-4-31B-it', 'Qwen3-32B', 'Intern-S1', 'MiniMax-M2', 'gemma-4-26B-A4B-it', 'Qwen3.5-4B', 'Olmo-3-32B-Think', 'Olmo-3-7B-Think', 'Llama-3.1-405B-Instruct-FP8', 'Llama-3.1-70B-Instruct', 'ChemDFM-R-14B', 'Llama-3.1-8B-Instruct' \\
Accuracy x Error-3 & 'Deepseek-v4', 'gpt-oss-120b', 'Qwen3.5-122B-A10B-FP8' \\
Accuracy x Error-5 & 'Deepseek-v4', 'gpt-oss-120b', 'Qwen3.5-122B-A10B-FP8', 'Qwen3-235B-A22B-Thinking-2507', 'Qwen3.5-27B' \\
Accuracy x Error-10 & 'Deepseek-v4', 'gpt-oss-120b', 'Qwen3.5-122B-A10B-FP8', 'Qwen3-235B-A22B-Thinking-2507', 'Qwen3.5-27B', 'gpt-oss-20b', 'Qwen3-Next-80B-A3B-Thinking', 'Qwen3.5-9B', 'MiniMax-M2', 'Qwen3.5-4B' \\
Accuracy x Error-15 & 'Deepseek-v4', 'gpt-oss-120b', 'Qwen3.5-122B-A10B-FP8', 'Qwen3-235B-A22B-Thinking-2507', 'Qwen3.5-27B', 'gpt-oss-20b', 'Qwen3-Next-80B-A3B-Thinking', 'Qwen3.5-9B', 'MiniMax-M2', 'Qwen3.5-4B', 'gemma-4-31B-it', 'Qwen3-32B', 'Qwen3.5-2B', 'gemma-4-26B-A4B-it', 'Intern-S1' \\
Accuracy x Error-20 & 'Deepseek-v4', 'gpt-oss-120b', 'Qwen3.5-122B-A10B-FP8', 'Qwen3-235B-A22B-Thinking-2507', 'Qwen3.5-27B', 'gpt-oss-20b', 'Qwen3-Next-80B-A3B-Thinking', 'Qwen3.5-9B', 'MiniMax-M2', 'Qwen3.5-4B', 'gemma-4-31B-it', 'Qwen3-32B', 'Qwen3.5-2B', 'gemma-4-26B-A4B-it', 'Intern-S1', 'Olmo-3-32B-Think', 'Olmo-3-7B-Think', 'Chem-R-8B', 'Llama-3.1-405B-Instruct-FP8', 'cosmosage-v3.1' \\
Error-3 & 'Qwen3.5-2B', 'MiniMax-M2', 'Qwen3.5-27B' \\
Error-5 & 'Qwen3.5-2B', 'MiniMax-M2', 'Qwen3.5-27B', 'Chem-R-8B', 'cosmosage-v3.1' \\
Error-10 & 'Qwen3.5-2B', 'MiniMax-M2', 'Qwen3.5-27B', 'Chem-R-8B', 'cosmosage-v3.1', 'Llama-3.1-8B-Instruct', 'ChemDFM-R-14B', 'Olmo-3-7B-Think', 'Llama-3.1-70B-Instruct', 'Olmo-3-32B-Think' \\
Error-15 & 'Qwen3.5-2B', 'MiniMax-M2', 'Qwen3.5-27B', 'Chem-R-8B', 'cosmosage-v3.1', 'Llama-3.1-8B-Instruct', 'ChemDFM-R-14B', 'Olmo-3-7B-Think', 'Llama-3.1-70B-Instruct', 'Olmo-3-32B-Think', 'Llama-3.1-405B-Instruct-FP8', 'gpt-oss-20b', 'gemma-4-31B-it', 'Qwen3-32B', 'Qwen3-Next-80B-A3B-Thinking' \\
Error-20 & 'Qwen3.5-2B', 'MiniMax-M2', 'Qwen3.5-27B', 'Chem-R-8B', 'cosmosage-v3.1', 'Llama-3.1-8B-Instruct', 'ChemDFM-R-14B', 'Olmo-3-7B-Think', 'Llama-3.1-70B-Instruct', 'Olmo-3-32B-Think', 'Llama-3.1-405B-Instruct-FP8', 'gpt-oss-20b', 'gemma-4-31B-it', 'Qwen3-32B', 'Qwen3-Next-80B-A3B-Thinking', 'Intern-S1', 'Qwen3.5-4B', 'Qwen3.5-9B', 'Deepseek-v4', 'gemma-4-26B-A4B-it' \\
IoU-3 & 'Chem-R-8B', 'Qwen3.5-2B', 'Deepseek-v4' \\
IoU-5 & 'Chem-R-8B', 'Qwen3.5-2B', 'Deepseek-v4', 'cosmosage-v3.1', 'Llama-3.1-8B-Instruct' \\
IoU-10 & 'Chem-R-8B', 'Qwen3.5-2B', 'Deepseek-v4', 'cosmosage-v3.1', 'Llama-3.1-8B-Instruct', 'ChemDFM-R-14B', 'Llama-3.1-405B-Instruct-FP8', 'Olmo-3-7B-Think', 'Llama-3.1-70B-Instruct', 'MiniMax-M2' \\
IoU-15 & 'Chem-R-8B', 'Qwen3.5-2B', 'Deepseek-v4', 'cosmosage-v3.1', 'Llama-3.1-8B-Instruct', 'ChemDFM-R-14B', 'Llama-3.1-405B-Instruct-FP8', 'Olmo-3-7B-Think', 'Llama-3.1-70B-Instruct', 'MiniMax-M2', 'gemma-4-26B-A4B-it', 'Olmo-3-32B-Think', 'gpt-oss-20b', 'Qwen3-Next-80B-A3B-Thinking', 'Qwen3.5-4B' \\
IoU-20 & 'Chem-R-8B', 'Qwen3.5-2B', 'Deepseek-v4', 'cosmosage-v3.1', 'Llama-3.1-8B-Instruct', 'ChemDFM-R-14B', 'Llama-3.1-405B-Instruct-FP8', 'Olmo-3-7B-Think', 'Llama-3.1-70B-Instruct', 'MiniMax-M2', 'gemma-4-26B-A4B-it', 'Olmo-3-32B-Think', 'gpt-oss-20b', 'Qwen3-Next-80B-A3B-Thinking', 'Qwen3.5-4B', 'Intern-S1', 'Qwen3.5-27B', 'Qwen3.5-9B', 'Qwen3-32B', 'gemma-4-31B-it' \\
\bottomrule
\end{tabularx}
\caption{The models contained in each evaluated subset (Part 1)}
\label{tab:all_subsets1}
\end{table*}

\begin{table*}
\centering
\scriptsize
\begin{tabularx}{\textwidth}{lX}
\toprule
\textbf{Run-Name}  & \textbf{Model-List} \\
\midrule
LLMchosen-GPT5-3 & 'Deepseek-v4', 'gemma-4-31B-it', 'gpt-oss-120b' \\
LLMchosen-GPT5-5 & 'Deepseek-v4', 'Qwen3.5-122B-A10B-FP8', 'gemma-4-31B-it', 'Qwen3-235B-A22B-Thinking-2507', 'gpt-oss-120b' \\
LLMchosen-GPT5-10 & 'Deepseek-v4', 'Qwen3.5-122B-A10B-FP8', 'gemma-4-31B-it', 'gpt-oss-120b', 'Intern-S1', 'Qwen3-235B-A22B-Thinking-2507', 'MiniMax-M2', 'Qwen3.5-27B', 'Qwen3-Next-80B-A3B-Thinking', 'ChemDFM-R-14B' \\
LLMchosen-GPT5-15 & 'Deepseek-v4', 'Qwen3.5-122B-A10B-FP8', 'gemma-4-31B-it', 'gpt-oss-120b', 'Intern-S1', 'Qwen3-235B-A22B-Thinking-2507', 'Qwen3.5-27B', 'gemma-4-26B-A4B-it', 'ChemDFM-R-14B', 'Qwen3.5-9B', 'MiniMax-M2', 'Qwen3-Next-80B-A3B-Thinking', 'Chem-R-8B', 'gpt-oss-20b', 'Qwen3.5-4B' \\
LLMchosen-GPT5-20 & 'Deepseek-v4', 'Qwen3.5-122B-A10B-FP8', 'gemma-4-31B-it', 'gpt-oss-120b', 'Intern-S1', 'Qwen3.5-27B', 'Qwen3-235B-A22B-Thinking-2507', 'MiniMax-M2', 'gemma-4-26B-A4B-it', 'ChemDFM-R-14B', 'Qwen3-Next-80B-A3B-Thinking', 'Chem-R-8B', 'Qwen3.5-9B', 'gpt-oss-20b', 'Qwen3-32B', 'Olmo-3-32B-Think', 'Llama-3.1-405B-Instruct-FP8', 'Qwen3.5-4B', 'cosmosage-v3.1', 'Llama-3.1-70B-Instruct' \\
LRMS-all & 'Intern-S1', 'MiniMax-M2', 'Qwen3.5-122B-A10B-FP8', 'gpt-oss-120b', 'Qwen3-Next-80B-A3B-Thinking', 'Olmo-3-32B-Think', 'gemma-4-31B-it', 'Deepseek-v4' \\
LRMs-3 & 'Deepseek-v4', 'Intern-S1', 'MiniMax-M2' \\
LRMs-5 & 'Deepseek-v4', 'Intern-S1', 'MiniMax-M2', 'Qwen3.5-122B-A10B-FP8', 'gpt-oss-120b' \\
Random-0-3 & 'Qwen3.5-9B', 'Llama-3.1-8B-Instruct', 'Qwen3-235B-A22B-Thinking-2507' \\
Random-0-5 & 'Qwen3.5-9B', 'Llama-3.1-8B-Instruct', 'Qwen3-235B-A22B-Thinking-2507', 'MiniMax-M2', 'Qwen3-32B' \\
Random-0-10 & 'Qwen3.5-9B', 'Llama-3.1-8B-Instruct', 'Qwen3-235B-A22B-Thinking-2507', 'MiniMax-M2', 'Qwen3-32B', 'Qwen3.5-2B', 'gpt-oss-20b', 'Deepseek-v4', 'ChemDFM-R-14B', 'Intern-S1' \\
Random-0-15 & 'Qwen3.5-9B', 'Llama-3.1-8B-Instruct', 'Qwen3-235B-A22B-Thinking-2507', 'MiniMax-M2', 'Qwen3-32B', 'Qwen3.5-2B', 'gpt-oss-20b', 'Deepseek-v4', 'ChemDFM-R-14B', 'Intern-S1', 'Qwen3.5-122B-A10B-FP8', 'Chem-R-8B', 'Llama-3.1-405B-Instruct-FP8', 'cosmosage-v3.1', 'Qwen3.5-4B' \\
Random-0-20 & 'Qwen3.5-9B', 'Llama-3.1-8B-Instruct', 'Qwen3-235B-A22B-Thinking-2507', 'MiniMax-M2', 'Qwen3-32B', 'Qwen3.5-2B', 'gpt-oss-20b', 'Deepseek-v4', 'ChemDFM-R-14B', 'Intern-S1', 'Qwen3.5-122B-A10B-FP8', 'Chem-R-8B', 'Llama-3.1-405B-Instruct-FP8', 'cosmosage-v3.1', 'Qwen3.5-4B', 'Olmo-3-7B-Think', 'Olmo-3-32B-Think', 'gpt-oss-120b', 'Llama-3.1-70B-Instruct', 'Qwen3-Next-80B-A3B-Thinking' \\
Random-1-3 & 'ChemDFM-R-14B', 'Llama-3.1-405B-Instruct-FP8', 'Chem-R-8B' \\
Random-1-5 & 'ChemDFM-R-14B', 'Llama-3.1-405B-Instruct-FP8', 'Chem-R-8B', 'Qwen3.5-27B', 'Qwen3-32B' \\
Random-1-10 & 'ChemDFM-R-14B', 'Llama-3.1-405B-Instruct-FP8', 'Chem-R-8B', 'Qwen3.5-27B', 'Qwen3-32B', 'MiniMax-M2', 'Olmo-3-7B-Think', 'Qwen3.5-4B', 'Qwen3.5-9B', 'gpt-oss-20b' \\
Random-1-15 & 'ChemDFM-R-14B', 'Llama-3.1-405B-Instruct-FP8', 'Chem-R-8B', 'Qwen3.5-27B', 'Qwen3-32B', 'MiniMax-M2', 'Olmo-3-7B-Think', 'Qwen3.5-4B', 'Qwen3.5-9B', 'gpt-oss-20b', 'gemma-4-31B-it', 'Llama-3.1-8B-Instruct', 'gpt-oss-120b', 'Qwen3.5-2B', 'cosmosage-v3.1' \\
Random-1-20 & 'ChemDFM-R-14B', 'Llama-3.1-405B-Instruct-FP8', 'Chem-R-8B', 'Qwen3.5-27B', 'Qwen3-32B', 'MiniMax-M2', 'Olmo-3-7B-Think', 'Qwen3.5-4B', 'Qwen3.5-9B', 'gpt-oss-20b', 'gemma-4-31B-it', 'Llama-3.1-8B-Instruct', 'gpt-oss-120b', 'Qwen3.5-2B', 'cosmosage-v3.1', 'Qwen3-235B-A22B-Thinking-2507', 'Olmo-3-32B-Think', 'Intern-S1', 'Llama-3.1-70B-Instruct', 'Qwen3-Next-80B-A3B-Thinking' \\
Random-2-3 & 'Llama-3.1-405B-Instruct-FP8', 'Chem-R-8B', 'gpt-oss-20b' \\
Random-2-5 & 'Llama-3.1-405B-Instruct-FP8', 'Chem-R-8B', 'gpt-oss-20b', 'Llama-3.1-8B-Instruct', 'gemma-4-31B-it' \\
Random-2-10 & 'Llama-3.1-405B-Instruct-FP8', 'Chem-R-8B', 'gpt-oss-20b', 'Llama-3.1-8B-Instruct', 'gemma-4-31B-it', 'Llama-3.1-70B-Instruct', 'Qwen3.5-122B-A10B-FP8', 'Olmo-3-7B-Think', 'Intern-S1', 'Deepseek-v4' \\
Random-2-15 & 'Llama-3.1-405B-Instruct-FP8', 'Chem-R-8B', 'gpt-oss-20b', 'Llama-3.1-8B-Instruct', 'gemma-4-31B-it', 'Llama-3.1-70B-Instruct', 'Qwen3.5-122B-A10B-FP8', 'Olmo-3-7B-Think', 'Intern-S1', 'Deepseek-v4', 'Qwen3-Next-80B-A3B-Thinking', 'Qwen3-235B-A22B-Thinking-2507', 'ChemDFM-R-14B', 'Qwen3.5-27B', 'Qwen3.5-2B' \\
Random-2-20 & 'Llama-3.1-405B-Instruct-FP8', 'Chem-R-8B', 'gpt-oss-20b', 'Llama-3.1-8B-Instruct', 'gemma-4-31B-it', 'Llama-3.1-70B-Instruct', 'Qwen3.5-122B-A10B-FP8', 'Olmo-3-7B-Think', 'Intern-S1', 'Deepseek-v4', 'Qwen3-Next-80B-A3B-Thinking', 'Qwen3-235B-A22B-Thinking-2507', 'ChemDFM-R-14B', 'Qwen3.5-27B', 'Qwen3.5-2B', 'gpt-oss-120b', 'MiniMax-M2', 'Qwen3.5-4B', 'gemma-4-26B-A4B-it', 'Qwen3.5-9B' \\
Random-3-3 & 'Olmo-3-7B-Think', 'Qwen3.5-122B-A10B-FP8', 'Olmo-3-32B-Think' \\
Random-3-5 & 'Olmo-3-7B-Think', 'Qwen3.5-122B-A10B-FP8', 'Olmo-3-32B-Think', 'MiniMax-M2', 'Llama-3.1-70B-Instruct' \\
Random-3-10 & 'Olmo-3-7B-Think', 'Qwen3.5-122B-A10B-FP8', 'Olmo-3-32B-Think', 'MiniMax-M2', 'Llama-3.1-70B-Instruct', 'Qwen3.5-4B', 'Llama-3.1-8B-Instruct', 'gpt-oss-20b', 'gemma-4-31B-it', 'Qwen3.5-2B' \\
Random-3-15 & 'Olmo-3-7B-Think', 'Qwen3.5-122B-A10B-FP8', 'Olmo-3-32B-Think', 'MiniMax-M2', 'Llama-3.1-70B-Instruct', 'Qwen3.5-4B', 'Llama-3.1-8B-Instruct', 'gpt-oss-20b', 'gemma-4-31B-it', 'Qwen3.5-2B', 'Chem-R-8B', 'Qwen3.5-27B', 'gemma-4-26B-A4B-it', 'Llama-3.1-405B-Instruct-FP8', 'cosmosage-v3.1' \\
\bottomrule
\end{tabularx}
\caption{The models contained in each evaluated subset (Part 2)}
\label{tab:all_subsets2}
\end{table*}

\begin{table*}
\centering
\scriptsize
\begin{tabularx}{\textwidth}{lX}
\toprule
\textbf{Run-Name}  & \textbf{Model-List} \\
\midrule
Random-3-20 & 'Olmo-3-7B-Think', 'Qwen3.5-122B-A10B-FP8', 'Olmo-3-32B-Think', 'MiniMax-M2', 'Llama-3.1-70B-Instruct', 'Qwen3.5-4B', 'Llama-3.1-8B-Instruct', 'gpt-oss-20b', 'gemma-4-31B-it', 'Qwen3.5-2B', 'Chem-R-8B', 'Qwen3.5-27B', 'gemma-4-26B-A4B-it', 'Llama-3.1-405B-Instruct-FP8', 'cosmosage-v3.1', 'Qwen3-235B-A22B-Thinking-2507', 'Qwen3-Next-80B-A3B-Thinking', 'gpt-oss-120b', 'ChemDFM-R-14B', 'Qwen3.5-9B' \\
Random-4-3 & 'Qwen3-32B', 'Llama-3.1-405B-Instruct-FP8', 'gpt-oss-120b' \\
Random-4-5 & 'Qwen3-32B', 'Llama-3.1-405B-Instruct-FP8', 'gpt-oss-120b', 'Intern-S1', 'Qwen3.5-122B-A10B-FP8' \\
Random-4-10 & 'Qwen3-32B', 'Llama-3.1-405B-Instruct-FP8', 'gpt-oss-120b', 'Intern-S1', 'Qwen3.5-122B-A10B-FP8', 'Qwen3-Next-80B-A3B-Thinking', 'Llama-3.1-8B-Instruct', 'Deepseek-v4', 'Qwen3.5-4B', 'Qwen3-235B-A22B-Thinking-2507' \\
Random-4-15 & 'Qwen3-32B', 'Llama-3.1-405B-Instruct-FP8', 'gpt-oss-120b', 'Intern-S1', 'Qwen3.5-122B-A10B-FP8', 'Qwen3-Next-80B-A3B-Thinking', 'Llama-3.1-8B-Instruct', 'Deepseek-v4', 'Qwen3.5-4B', 'Qwen3-235B-A22B-Thinking-2507', 'Llama-3.1-70B-Instruct', 'Olmo-3-32B-Think', 'Chem-R-8B', 'gemma-4-26B-A4B-it', 'gemma-4-31B-it' \\
Random-4-20 & 'Qwen3-32B', 'Llama-3.1-405B-Instruct-FP8', 'gpt-oss-120b', 'Intern-S1', 'Qwen3.5-122B-A10B-FP8', 'Qwen3-Next-80B-A3B-Thinking', 'Llama-3.1-8B-Instruct', 'Deepseek-v4', 'Qwen3.5-4B', 'Qwen3-235B-A22B-Thinking-2507', 'Llama-3.1-70B-Instruct', 'Olmo-3-32B-Think', 'Chem-R-8B', 'gemma-4-26B-A4B-it', 'gemma-4-31B-it', 'Olmo-3-7B-Think', 'Qwen3.5-9B', 'MiniMax-M2', 'Qwen3.5-2B', 'gpt-oss-20b' \\
Shared-Gemma & 'gemma-4-26B-A4B-it', 'gemma-4-31B-it' \\
Shared-Llama  & 'Llama-3.1-8B-Instruct', 'Llama-3.1-70B-Instruct', 'Llama-3.1-405B-Instruct-FP8', 'cosmosage-v3.1', 'Chem-R-8B' \\
Shared-Olmo & 'Olmo-3-32B-Think', 'Olmo-3-7B-Think' \\
Shared-Qwen3 & 'Qwen3-235B-A22B-Thinking-2507', 'Intern-S1', 'Qwen3-32B' \\
Shared-Qwen3.5 & 'Qwen3.5-122B-A10B-FP8', 'Qwen3.5-27B', 'Qwen3.5-9B', 'Qwen3.5-4B', 'Qwen3.5-2B' \\
Shared-gpt & 'gpt-oss-20b', 'gpt-oss-120b' \\
\bottomrule
\end{tabularx}
\caption{The models contained in each evaluated subset (Part 3)}
\label{tab:all_subsets3}
\end{table*}

\section{Results on all datasets}
\label{app:all}

\subsection{Oracle Performance}\label{app:oracle}

\begin{figure*}
    \centering
    \includegraphics[width=\linewidth]{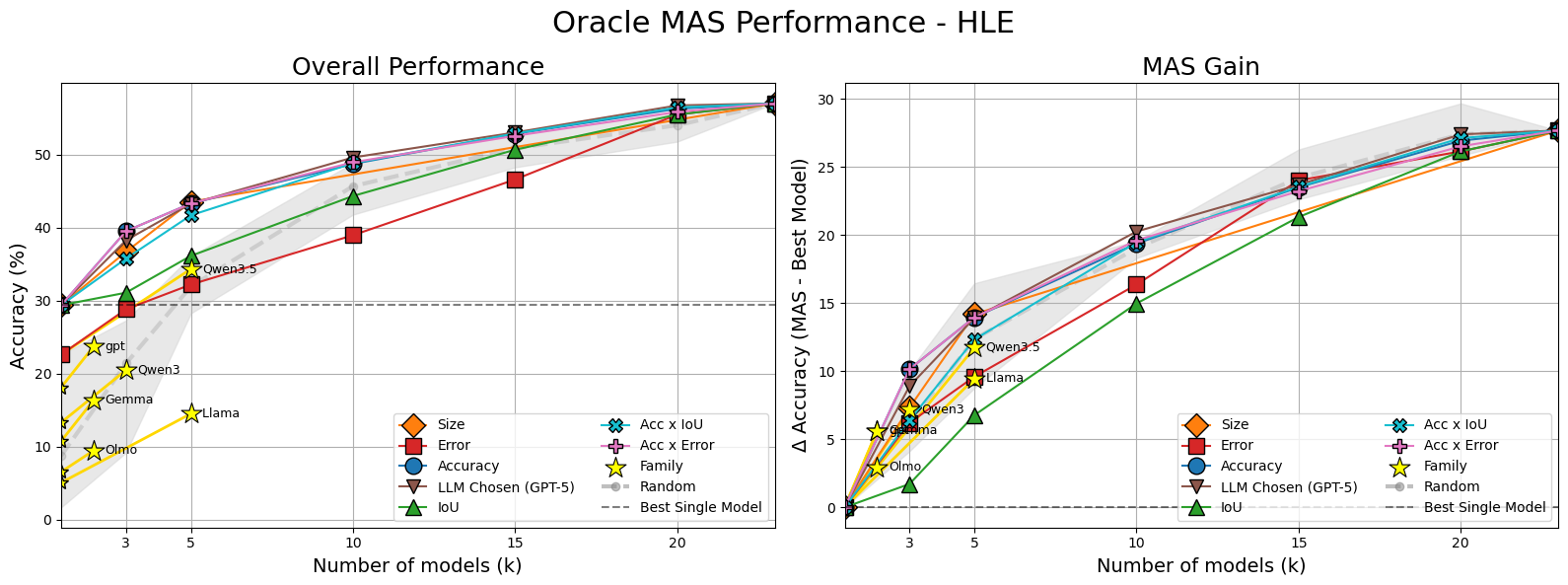}
    \caption{Oracle performances of each $C_{s,k}$ with increasing $k$ on HLE}
    \label{fig:oracle_hle}
\end{figure*}

\begin{figure*}
    \centering
    \includegraphics[width=\linewidth]{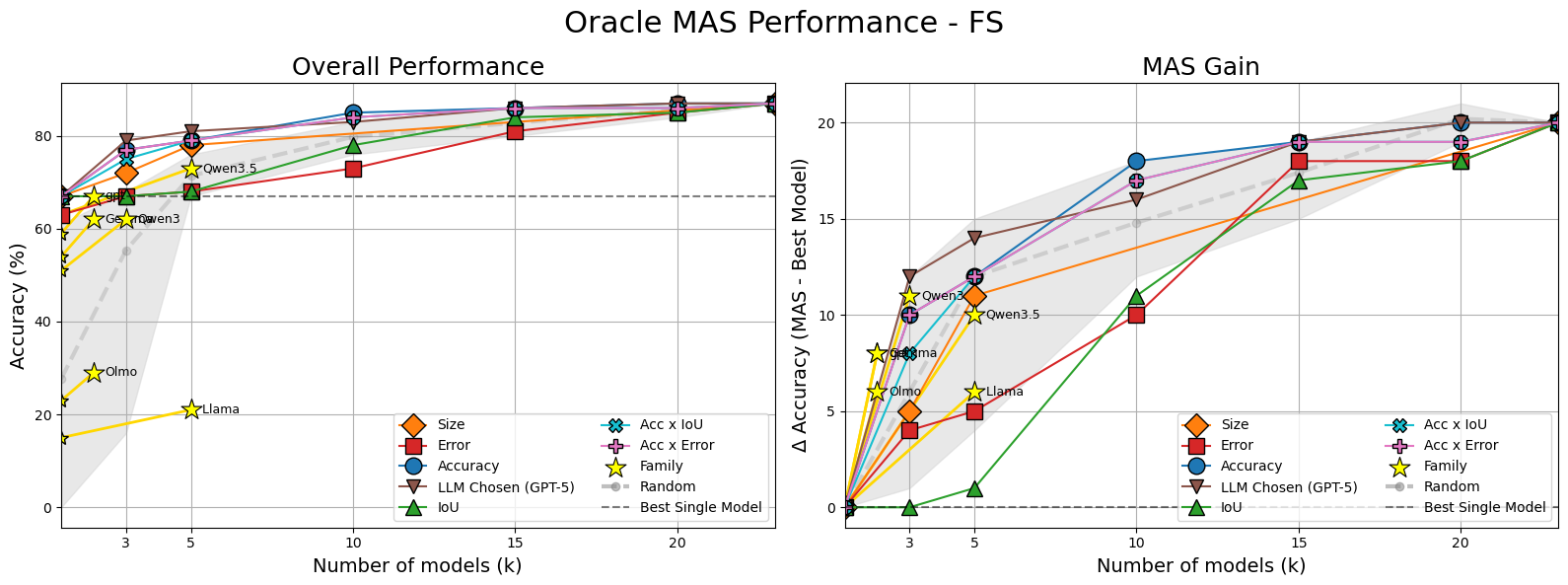}
    \caption{Oracle performances of each $C_{s,k}$ with increasing $k$ on FrontierScience-Olympiad}
    \label{fig:oracle_fs}
\end{figure*}

\begin{figure*}
    \centering
    \includegraphics[width=\linewidth]{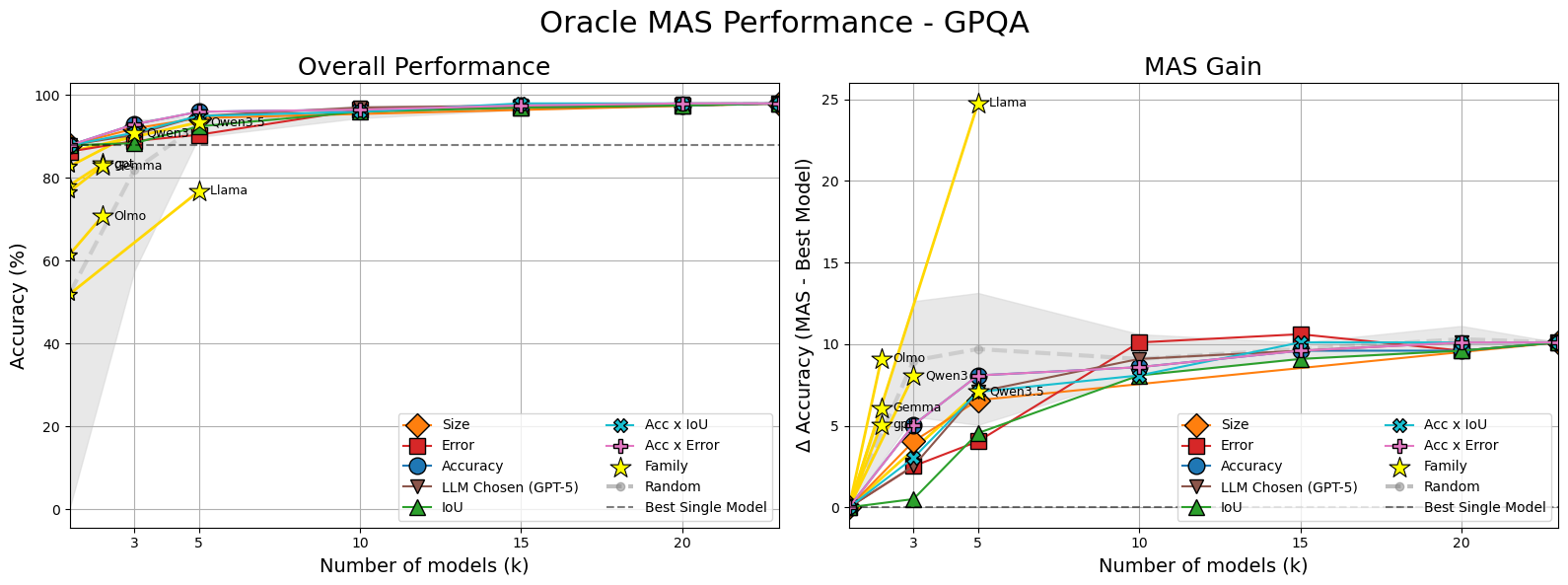}
    \caption{Oracle performances of each $C_{s,k}$ with increasing $k$ on GPQA-Diamond}
    \label{fig:oracle_gpqa}
\end{figure*}

We show the oracle performance of each MAS on HLE in \Cref{fig:oracle_hle}, on FS in \Cref{fig:oracle_fs}, and on GPQA in \Cref{fig:oracle_gpqa}. On average, performance is highest on GPQA, followed closely by FS. Given the low overhead available on GPQA for improvement, we see the greatest prophesized improvements on HLE and FS. Typically, we see the greatest predicted MAS gains when grouping by Accuracy \Acc \hspace{3pt} or LLM suggestions \Llm \hspace{3pt}. Typically, we see limited prophesized improvements when grouping by shared architecture \Fam \hspace{3pt}; once exception is on GPQA with the Llama family. As this is the easiest science reasoning dataset, this may arise from successful responses from our specialist Llama-based models.

\subsection{Actual Performance}\label{app:actual}

\begin{figure*}[t]
    \centering
    \includegraphics[width=\linewidth]{images/hle_actual.png}
    \caption{Actual MAS Gain of each $C_{s,k}$ given three different MAS architectures on Humanity's Last Exam}
    \label{fig:actual_hle}

    \vspace{1em}

    \includegraphics[width=\linewidth]{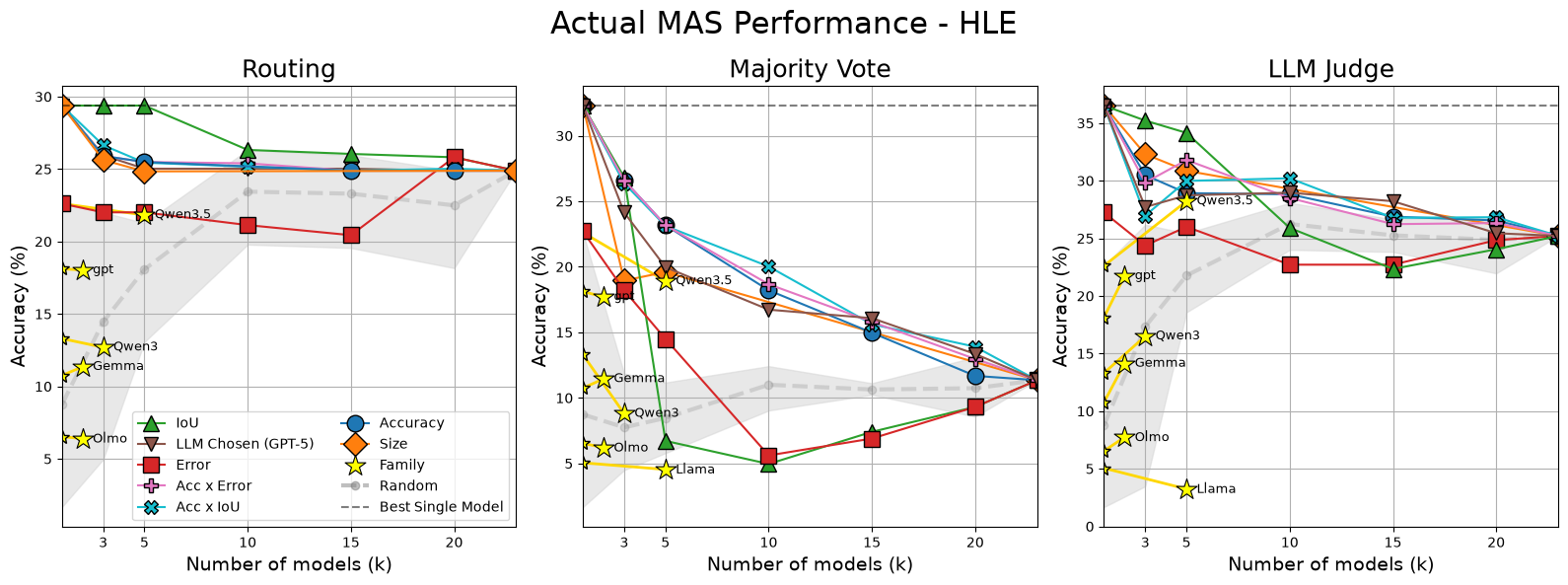}
    \caption{Actual performances (in Accuracy) of each $C_{s,k}$ given three different MAS architectures on Humanity's Last Exam}
    \label{fig:actual_hle_acc}
\end{figure*}

\begin{figure*}[t]
    \centering
    \includegraphics[width=\linewidth]{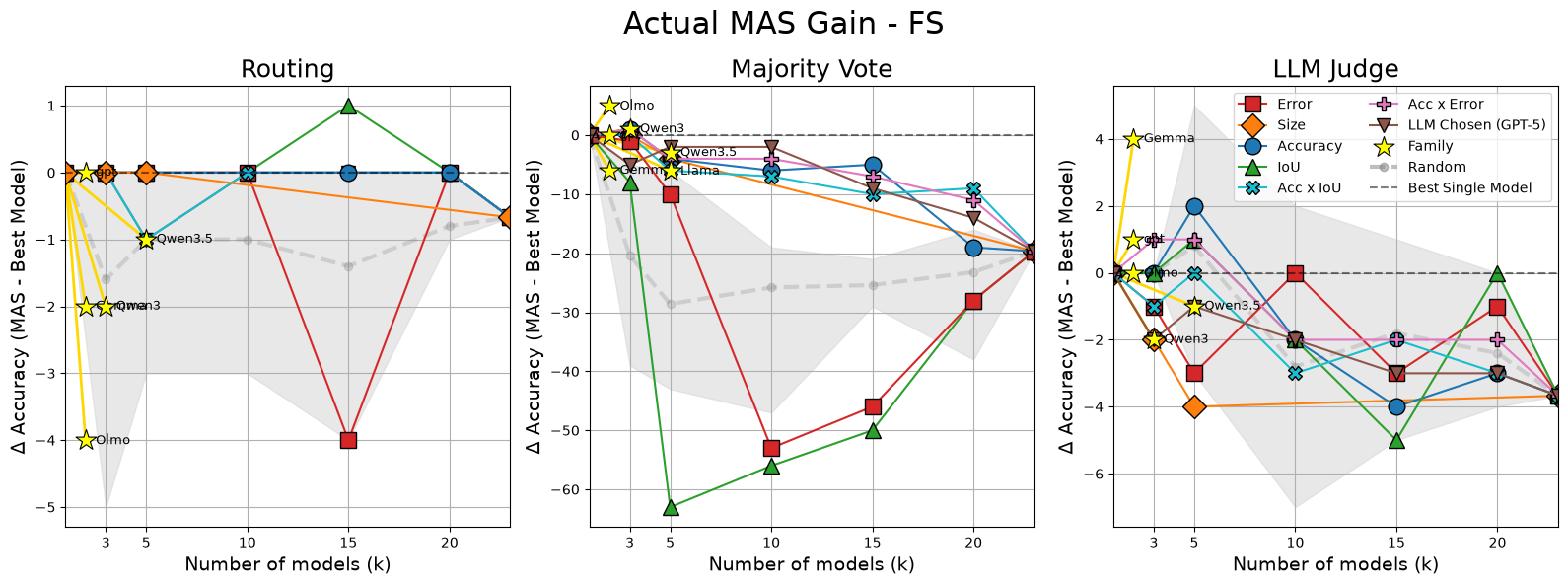}
    \caption{Actual MAS Gain of each $C_{s,k}$ given three different MAS architectures on FrontierScience-Olympiad}
    \label{fig:actual_fs}
    
    \vspace{1em}

    \includegraphics[width=\linewidth]{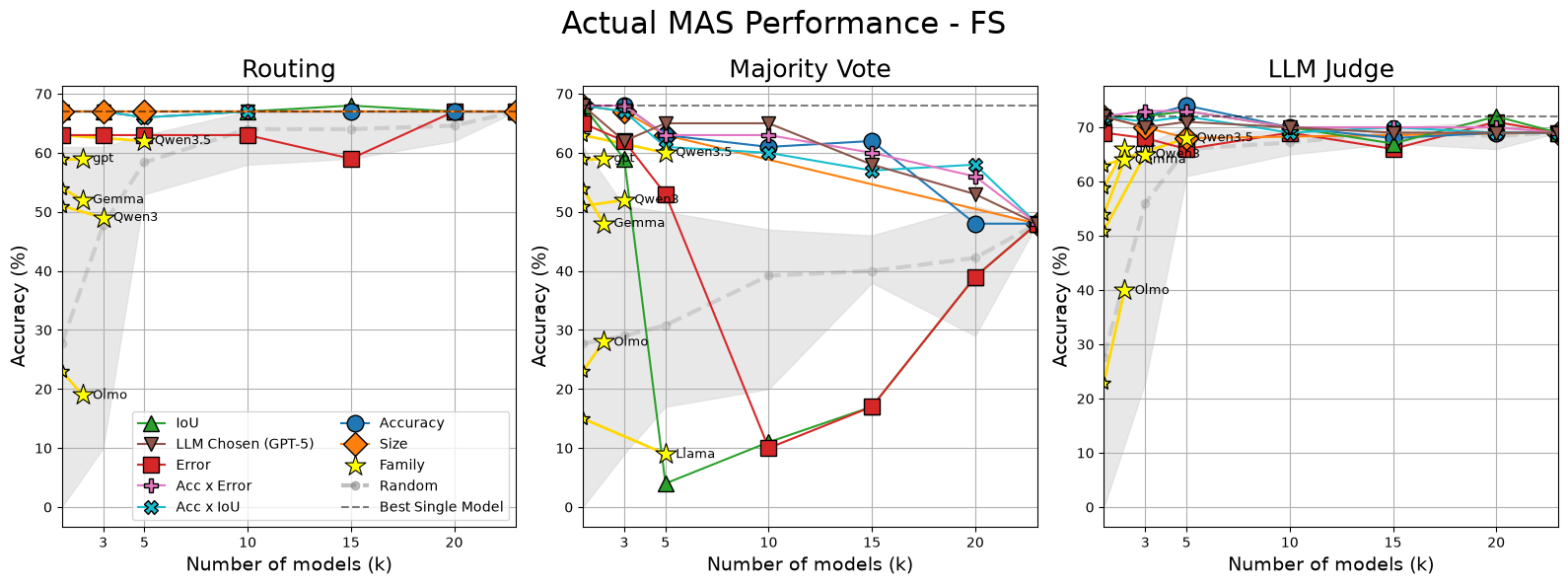}
    \caption{Actual performances (in Accuracy) of each $C_{s,k}$ given three different MAS architectures on FrontierScience-Olympiad}
    \label{fig:actual_fs_acc}
\end{figure*}

\begin{figure*}[t]
    \centering
    \includegraphics[width=\linewidth]{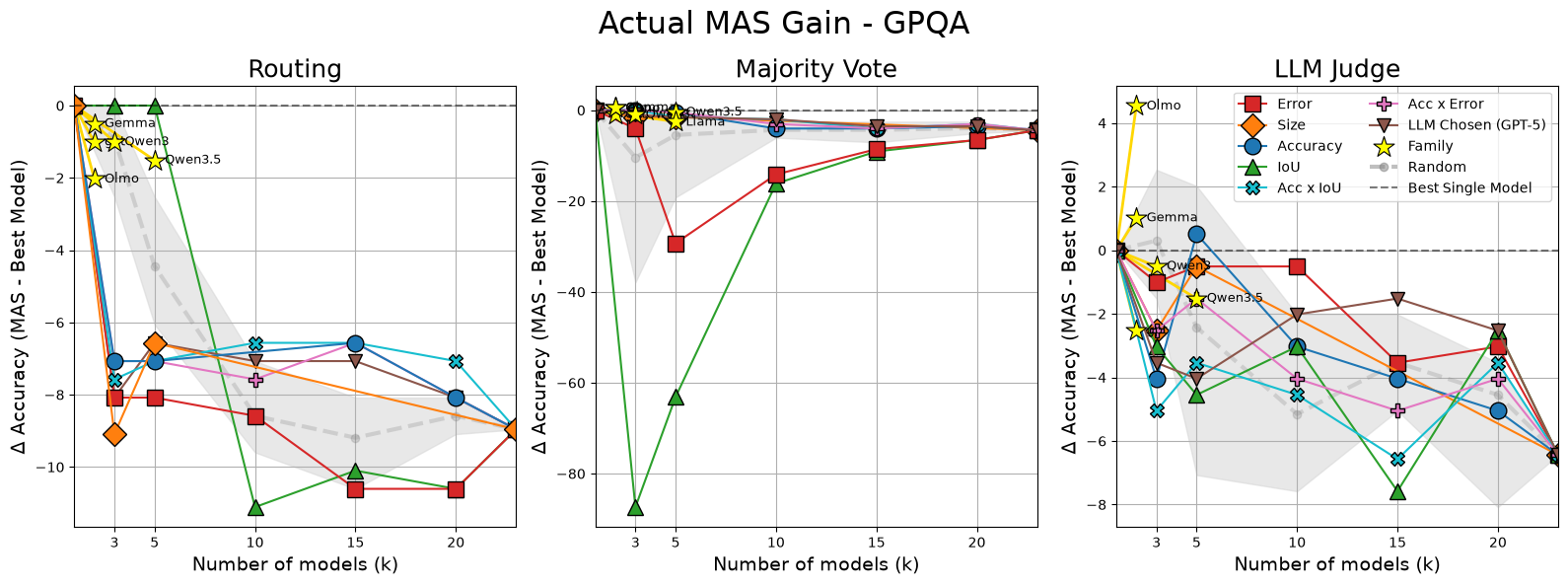}
    \caption{Actual MAS Gain of each $C_{s,k}$ given three different MAS architectures on GPQA-Diamond}
    \label{fig:actual_gpqa}
    
    \vspace{1em}

    \includegraphics[width=\linewidth]{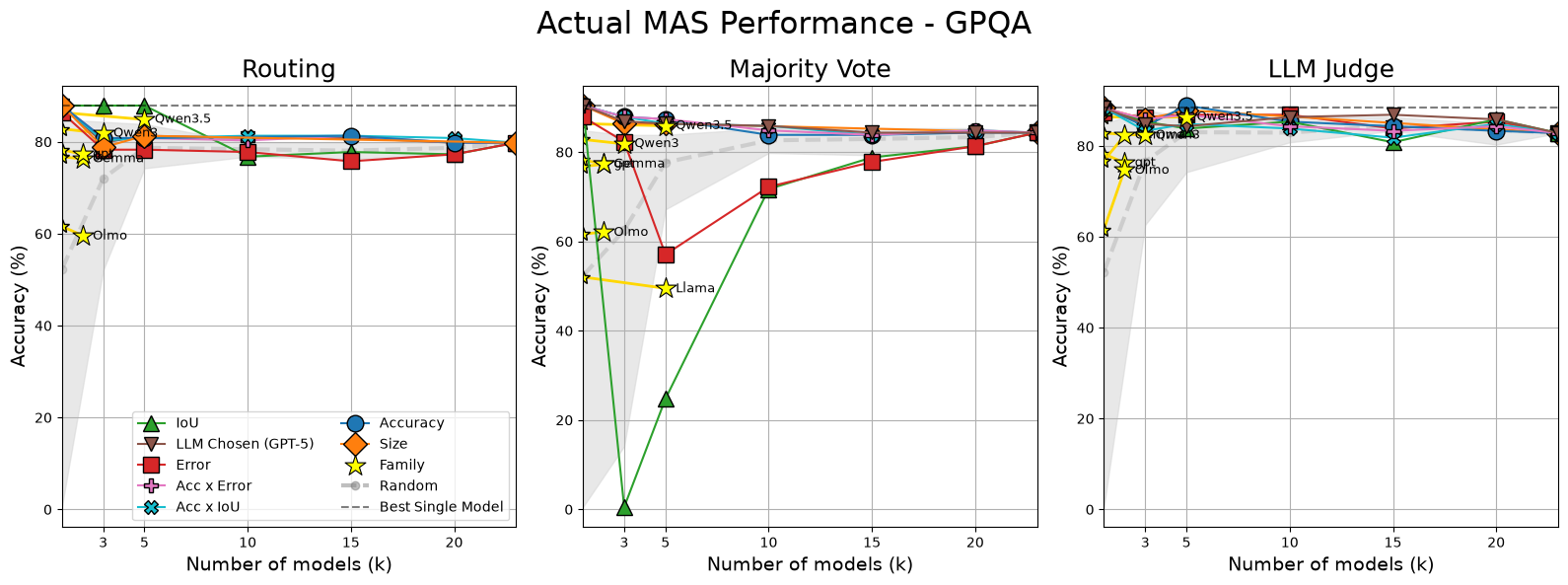}
    \caption{Actual performances (in Accuracy) of each $C_{s,k}$ given three different MAS architectures on GPQA-Diamond}
    \label{fig:actual_gpqa_acc}
\end{figure*}

We show the actual performance of each MAS on HLE in \Cref{fig:actual_hle,fig:actual_hle_acc}, on FS in \Cref{fig:actual_fs,fig:actual_fs_acc}, and on GPQA in \Cref{fig:actual_gpqa,fig:actual_gpqa_acc}. We see the steepest decrease in performance with increasing $k$ on HLE. MAS built from the gemma4 family of models showed the best performance. When we look at FS, we see a limited drop in our routing system performance, and we actually see an improvement above the base model for $C_{IoU,15}$. The OLMo3 family \Fam \hspace{3pt} shows an improvement in performance in the majority-vote system, and IoU \IoU \hspace{3pt} and Error \Err \hspace{3pt} based grouping shows steep decreases in performance (as these groupings optimise for response diversity, it makes sense it would be difficult to get a robust majority response). LLM-as-a-Judge MAS groups identified using accuracy-based metrics (\Acc \AxI \AxE) also show an improvement over base-model performance on FS. Performance on GPQA follows a similar pattern for the other two models: prioritising IoU \IoU\hspace{3pt} has a limited negative effect on routing systems at low k, but has a strong detrimental impact on majority-vote MAS. Like on FS, we also see an improvement over our base model with $C_{Acc,5}$ for LLM-as-a-Judge MAS on GPQA. Overall, we see a starker difference in performance when we look at MAS gain, rather than accuracy. However, we still see drops in performance with increasing levels of $k$.

\end{document}